\documentclass[preprint]{aastex701}

\usepackage{amsmath}
\usepackage{soul}
\usepackage{xspace}

\begin{document}

\title{Measurement of the  photosphere oblateness
of $\gamma$ Cassiopeiae via Stellar Intensity Interferometry with the VERITAS Observatory}

\correspondingauthor{Josephine Rose}

\author[orcid=0000-0002-8967-3222]{A.~Archer}
\email{averyarcher@depauw.edu}
\affiliation{Department of Physics and Astronomy, DePauw University, Greencastle, IN 46135-0037, USA}

\author[orcid=0000-0002-4104-7580]{J.~P.~Aufdenberg}
\email{aufded93@erau.edu}
\affiliation{Embry-Riddle Aeronautical University, Physical Sciences Department, 1 Aerospace Blvd, Daytona Beach, FL 32114, USA}

\author[orcid=0000-0002-3886-3739]{P.~Bangale}
\email{pbangale@udel.edu}
\affiliation{Department of Physics and Astronomy and the Bartol Research Institute, University of Delaware, Newark, DE 19716, USA}

\author[orcid=0000-0002-9675-7328]{J.~T.~Bartkoske}
\email{joshua.bartkoske@utah.edu}
\affiliation{Department of Physics and Astronomy, University of Utah, Salt Lake City, UT 84112, USA}

\author[orcid=0000-0003-2098-170X]{W.~Benbow}
\email{wbenbow@cfa.harvard.edu}
\affiliation{Center for Astrophysics $|$ Harvard \& Smithsonian, Cambridge, MA 02138, USA}

\author[orcid=0000-0001-6391-9661]{J.~H.~Buckley}
\email{buckley@wustl.edu}
\affiliation{Department of Physics, Washington University, St. Louis, MO 63130, USA}

\author[orcid=0009-0001-5719-936X]{Y.~Chen}
\email{ychen@astro.ucla.edu}
\affiliation{Department of Physics and Astronomy, University of California, Los Angeles, CA 90095, USA}

\author[orcid=0009-0000-1447-8353]{N.~B.~Y.~Chin}
\email{u1552776@utah.edu}
\affiliation{University of Utah, Department of Physics and Astronomy, 115 South 1400 East 201, Salt Lake City, Utah, USA}

\author[orcid=0000-0001-5811-9678]{J.~L.~Christiansen}
\email{jlchrist@calpoly.edu}
\affiliation{Physics Department, California Polytechnic State University, San Luis Obispo, CA 94307, USA}

\author{A.~J.~Chromey}
\email{alisha.chromey@cfa.harvard.edu}
\affiliation{Center for Astrophysics $|$ Harvard \& Smithsonian, Cambridge, MA 02138, USA}

\author[orcid=0000-0003-1716-4119]{A.~Duerr}
\email{a.duerr@utah.edu}
\affiliation{Department of Physics and Astronomy, University of Utah, Salt Lake City, UT 84112, USA}

\author{M.~Escobar~Godoy}
\email{mescob11@ucsc.edu}
\affiliation{Santa Cruz Institute for Particle Physics and Department of Physics, University of California, Santa Cruz, CA 95064, USA}

\author[orcid=0000-0002-4234-1933]{S.~Feldman}
\email{sydneyfeldman@physics.ucla.edu}
\affiliation{Department of Physics and Astronomy, University of California, Los Angeles, CA 90095, USA}

\author[orcid=0000-0001-6674-4238]{Q.~Feng}
\email{qi.feng@utah.edu}
\affiliation{Department of Physics and Astronomy, University of Utah, Salt Lake City, UT 84112, USA}

\author[orcid=0000-0002-2636-4756]{S.~Filbert}
\email{u1477296@utah.edu}
\affiliation{Department of Physics and Astronomy, University of Utah, Salt Lake City, UT 84112, USA}

\author[orcid=0000-0002-1067-8558]{L.~Fortson}
\email{lffortson@gmail.com}
\affiliation{School of Physics and Astronomy, University of Minnesota, Minneapolis, MN 55455, USA}

\author[orcid=0000-0003-1614-1273]{A.~Furniss}
\email{afurniss@ucsc.edu}
\affiliation{Santa Cruz Institute for Particle Physics and Department of Physics, University of California, Santa Cruz, CA 95064, USA}

\author[orcid=0000-0002-0109-4737]{W.~Hanlon}
\email{william.hanlon@cfa.harvard.edu}
\affiliation{Center for Astrophysics $|$ Harvard \& Smithsonian, Cambridge, MA 02138, USA}

\author[orcid=0000-0003-3878-1677]{O.~Hervet}
\email{ohervet@ucsc.edu}
\affiliation{Santa Cruz Institute for Particle Physics and Department of Physics, University of California, Santa Cruz, CA 95064, USA}

\author[orcid=0000-0001-6951-2299]{C.~E.~Hinrichs}
\email{claire.hinrichs@cfa.harvard.edu}
\affiliation{Center for Astrophysics $|$ Harvard \& Smithsonian, Cambridge, MA 02138, USA}
\affiliation{Department of Physics and Astronomy, Dartmouth College, 6127 Wilder Laboratory, Hanover, NH 03755 USA}

\author[orcid=0000-0002-6833-0474]{J.~Holder}
\email{jholder@udel.edu}
\affiliation{Department of Physics and Astronomy and the Bartol Research Institute, University of Delaware, Newark, DE 19716, USA}

\author{Z.~Hughes}
\email{zdhughes@wustl.edu}
\affiliation{Department of Physics, Washington University, St. Louis, MO 63130, USA}

\author[orcid=0000-0002-1432-7771]{T.~B.~Humensky}
\email{brian.humensky@nasa.gov}
\affiliation{Department of Physics, University of Maryland, College Park, MD, USA}
\affiliation{NASA GSFC, Greenbelt, MD 20771, USA}

\author[orcid=0000-0002-1089-1754]{W.~Jin}
\email{wjin@astro.ucla.edu}
\affiliation{Department of Physics and Astronomy, University of California, Los Angeles, CA 90095, USA}

\author[orcid=0009-0008-2688-0815]{M.~N.~Johnson}
\email{mjohns56@ucsc.edu}
\affiliation{Santa Cruz Institute for Particle Physics and Department of Physics, University of California, Santa Cruz, CA 95064, USA}

\author{M.~Kertzman}
\email{kertzman@depauw.edu}
\affiliation{Department of Physics and Astronomy, DePauw University, Greencastle, IN 46135-0037, USA}

\author[orcid=0000-0003-4686-0922]{M.~Kherlakian}
\email{maria.kherlakian@astro.ruhr-uni-bochum.de}
\affiliation{Fakult\"at f\"ur Physik \& Astronomie, Ruhr-Universit\"at Bochum, D-44780 Bochum, Germany}

\author[orcid=0000-0003-4785-0101]{D.~Kieda}
\email{dave.kieda@utah.edu}
\affiliation{University of Utah, Department of Physics and Astronomy, 115 South 1400 East 201, Salt Lake City, Utah, USA}

\author[orcid=0000-0002-4289-7106]{N.~Korzoun}
\email{nkorzoun@udel.edu}
\affiliation{Department of Physics and Astronomy and the Bartol Research Institute, University of Delaware, Newark, DE 19716, USA}

\author[orcid=0000-0002-3489-7325]{T.~LeBohec}
\email{tugdual.lebohec@gmail.com}
\affiliation{University of Utah, Department of Physics and Astronomy, 115 South 1400 East 201, Salt Lake City, Utah, USA}

\author[orcid=0000-0001-5047-5213]{M.~A.~Lisa}
\email{lisa@physics.osu.edu}
\affiliation{The Ohio State University, Department of Physics, 191 W Woodruff Ave, Columbus, Ohio, USA}

\author[orcid=0000-0003-3802-1619]{M.~Lundy}
\email{matthew.lundy@mail.mcgill.ca}
\affiliation{Physics Department, McGill University, Montreal, QC H3A 2T8, Canada}

\author[orcid=0000-0001-9868-4700]{G.~Maier}
\email{gernot.maier@desy.de}
\affiliation{DESY, Platanenallee 6, 15738 Zeuthen, Germany}

\author[orcid=0000-0002-3687-4661]{N.~Matthews}
\email{nolankmatthews@gmail.com}
\affiliation{Space Dynamics Laboratory, Utah State University, Logan, UT, USA}
\affiliation{Université Côte d'Azur, CNRS, Institut de Physique de Nice, France}

\author[orcid=0000-0002-1499-2667]{P.~Moriarty}
\email{pmoriarty0@gmail.com}
\affiliation{School of Natural Sciences, University of Galway, University Road, Galway, H91 TK33, Ireland}

\author[orcid=0000-0002-3223-0754]{R.~Mukherjee}
\email{rmukherj@barnard.edu}
\affiliation{Department of Physics and Astronomy, Barnard College, Columbia University, NY 10027, USA}

\author[orcid=0000-0002-6121-3443]{W.~Ning}
\email{wning@astro.ucla.edu}
\affiliation{Department of Physics and Astronomy, University of California, Los Angeles, CA 90095, USA}

\author[orcid=0000-0002-4837-5253]{R.~A.~Ong}
\email{rene@astro.ucla.edu}
\affiliation{Department of Physics and Astronomy, University of California, Los Angeles, CA 90095, USA}

\author[orcid=0000-0003-3820-0887]{A.~Pandey}
\email{u6059187@utah.edu}
\affiliation{Department of Physics and Astronomy, University of Utah, Salt Lake City, UT 84112, USA}

\author[orcid=0000-0001-7861-1707]{M.~Pohl}
\email{martin.pohl@desy.de}
\affiliation{Institute of Physics and Astronomy, University of Potsdam, 14476 Potsdam-Golm, Germany}
\affiliation{DESY, Platanenallee 6, 15738 Zeuthen, Germany}

\author[orcid=0000-0002-0529-1973]{E.~Pueschel}
\email{elisa.pueschel@astro.ruhr-uni-bochum.de}
\affiliation{Fakult\"at f\"ur Physik \& Astronomie, Ruhr-Universit\"at Bochum, D-44780 Bochum, Germany}

\author[orcid=0000-0002-4855-2694]{J.~Quinn}
\email{john.quinn@ucd.ie}
\affiliation{School of Physics, University College Dublin, Belfield, Dublin 4, Ireland}

\author[orcid=0000-0002-5104-5263]{P.~L.~Rabinowitz}
\email{p.l.rabinowitz@wustl.edu}
\affiliation{Department of Physics, Washington University, St. Louis, MO 63130, USA}

\author[orcid=0000-0002-5351-3323]{K.~Ragan}
\email{ragan@physics.mcgill.ca}
\affiliation{Physics Department, McGill University, Montreal, QC H3A 2T8, Canada}

\author{P.~T.~Reynolds}
\email{Josh.Reynolds@mtu.ie}
\affiliation{Department of Physical Sciences, Munster Technological University, Bishopstown, Cork, T12 P928, Ireland}

\author[orcid=0000-0002-7523-7366]{D.~Ribeiro}
\email{ribei056@umn.edu}
\affiliation{School of Physics and Astronomy, University of Minnesota, Minneapolis, MN 55455, USA}

\author{E.~Roache}
\email{eroache@cfa.harvard.edu}
\affiliation{Center for Astrophysics $|$ Harvard \& Smithsonian, Cambridge, MA 02138, USA}

\author[orcid=0009-0000-0003-6407]{J.~G.~Rose}
\email[show]{josieg.rose.86@gmail.com}
\affiliation{The Ohio State University, Department of Physics, 191 W Woodruff Ave, Columbus, Ohio, USA}

\author[0000-0003-1387-8915]{I.~Sadeh}
\email{iftach.sadeh@desy.de}
\affiliation{DESY, Platanenallee 6, 15738 Zeuthen, Germany}

\author[orcid=0000-0002-3171-5039]{L.~Saha}
\email{lab.saha@cfa.harvard.edu}
\affiliation{Center for Astrophysics $|$ Harvard \& Smithsonian, Cambridge, MA 02138, USA}

\author[orcid=0000-0001-7297-8217]{M.~Santander}
\email{jmsantander@ua.edu}
\affiliation{Department of Physics and Astronomy, University of Alabama, Tuscaloosa, AL 35487, USA}

\author[orcid=0009-0005-9604-5127]{J.~Scott}
\email{scott.2752@osu.edu}
\affiliation{The Ohio State University, Department of Physics, 191 W Woodruff Ave, Columbus, Ohio, USA}

\author{G.~H.~Sembroski}
\email{sembrosk@purdue.edu}
\affiliation{Department of Physics and Astronomy, Purdue University, West Lafayette, IN 47907, USA}

\author[orcid=0000-0002-9856-989X]{R.~Shang}
\email{r.y.shang@gmail.com}
\affiliation{Department of Physics and Astronomy, Barnard College, Columbia University, NY 10027, USA}

\author[orcid=0000-0002-9852-2469]{D.~Tak}
\email{donggeun.tak@gmail.com}
\affiliation{SNU Astronomy Research Center, Seoul National University, Seoul 08826, Republic of Korea.}

\author{J.~V.~Tucci}
\email{jtucci@iu.edu}
\affiliation{Department of Physics, Indiana University Indianapolis, Indianapolis, Indiana 46202, USA}

\author[orcid=0000-0002-8090-6528]{J.~Valverde}
\email{janeth.valverdequispe@nasa.gov}
\affiliation{Department of Physics, University of Maryland, Baltimore County, Baltimore MD 21250, USA and NASA GSFC, Greenbelt, MD 20771, USA}

\author{V.~V.~Vassiliev}
\email{vvv@physics.ucla.edu}
\affiliation{Department of Physics and Astronomy, University of California, Los Angeles, CA 90095, USA}

\author[0000-0003-2740-9714]{D.~A.~Williams}
\email{daw@ucsc.edu}
\affiliation{Santa Cruz Institute for Particle Physics and Department of Physics, University of California, Santa Cruz, CA 95064, USA}

\author[orcid=0000-0002-2730-2733]{S.~L.~Wong}
\email{samantha.wong2@mail.mcgill.ca}
\affiliation{Physics Department, McGill University, Montreal, QC H3A 2T8, Canada}

\collaboration{60}{The VERITAS Collaboration}

\begin{abstract}
\nolinenumbers
We use the stellar intensity interferometry system implemented with the 
Very Energetic Radiation Imaging Telescope Array System (VERITAS)
at Fred Lawrence Whipple Observatory (FLWO) as a light collector to obtain measurements of the rapid rotator star $\gamma$ Cassiopeiae, at a
wavelength of $416\,\rm nm$. 
Using data from baselines sampling different position angles, we extract
the size, oblateness, and projected orientation of the photosphere.
Fitting the data with a uniform ellipse model yields a minor-axis angular diameter of
$0.43\pm0.02$\,mas, a major-to-minor-radius ratio of $1.28\pm0.04$, and a position
angle of $116^\circ\pm5^\circ$ for the axis of rotation.
A rapidly-rotating stellar atmosphere model that includes limb and gravity darkening describes the
data well with a fitted angular diameter of 
$0.604^{+0.041}_{-0.034}$~mas corresponding to an
equatorial radius of 10.9$^{+0.8}_{-0.6}~R_\odot$, a rotational
velocity with a $1~\sigma$ lower limit at $97.7\%$ that of breakup velocity, and a position angle of $114.7^{+6.4}_{-5.7}$ degrees. These parameters are consistent with 
H$\alpha$ line spectroscopy
and infrared-wavelength Michelson interferometric measurements of the star's
decretion disk.
This is the first measurement of an oblate photosphere using intensity interferometry.
\end{abstract}


\section{Introduction} \label{sec:Intro}

Spectroscopic measurements of the H$\alpha$ line of the subgiant $\gamma$ Cassiopeiae ($\gamma$ Cas) indicate that it is a Be-type star rotating at nearly its breakup
velocity about an axis inclined by $60^\circ$ to our line of sight.
Thus, it is expected to have a pronounced equatorial bulge.
The history of Michelson interferometric measurements of $\gamma$ Cas spans many decades~\citep[see \S 3 of][for a brief overview]{refId0}.
These measurements reveal an extended decretion disk that 
suggests 
a $122^\circ$ position angle of the rotational axis, measured east of celestial north~\citep{sigut_2020}.
However, these measurements, performed at infrared wavelengths, are unable to measure the photosphere itself.

Stellar intensity interferometry (SII) at optical wavelengths has
the potential to measure the photosphere itself at sub-milliarcsecond
scale.
Originally developed in the 1950's and used through the early 1970's~\citep{Brown01071954,1974RHB_book},
SII has very recently seen a 
renaissance with the availability of large light collectors in
imaging atmospheric Cherenkov telescope 
(IACT) arrays, high-speed electronics and modern computing at several observatories~\citep{natureSIIDemo,10.1093/mnras/stae697,10.1093/mnras/stad3676,VERITAS:2024quv}.
By correlating the intensity of light at
separated telescopes, rather than the wave amplitude
(as done with Michelson interferometry),  SII is robust against effects of optical imperfections, including atmospheric turbulence, and works well at optical wavelengths, providing sub-milliarcsecond resolutions with 100\,m baselines.
Relative to Michelson interferometry, this comes at a cost of reduced
signal-to-noise for the same observing duration.

In this paper, we present the first SII measurement sensitive to the
shape, as well as the size, of a star.
The work is organized as follows.
Section~\ref{sec:InstrumentAndDataset} gives a brief description of
the VERITAS stellar intensity interferometer (VSII) and the datasets discussed here.
Section~\ref{sec:Analysis} presents the data analysis and extracted
squared visibilities.
Section~\ref{sec:SimpleGeometricModelFit} uses simple geometric models of the 
star to extract size, shape, and orientation information from the measurements.
The dependence of the squared visibility on $u$ and $v$ is fitted
with uniform disk and uniform ellipse models.
Section~\ref{sec:Modeling} discusses a complete stellar model of a rapid
rotator with limb and gravitational darkening.
Comparison of our data with this model provides tight constraints on the size,
rotational speed, and orientation of the star.
In Section~\ref{sec:Summary}, we summarize and provide an outlook for
future work.

\section{Instrument and Dataset}
\label{sec:InstrumentAndDataset}

\begin{figure}[t]
    \centering
    \includegraphics[width=0.5\linewidth]{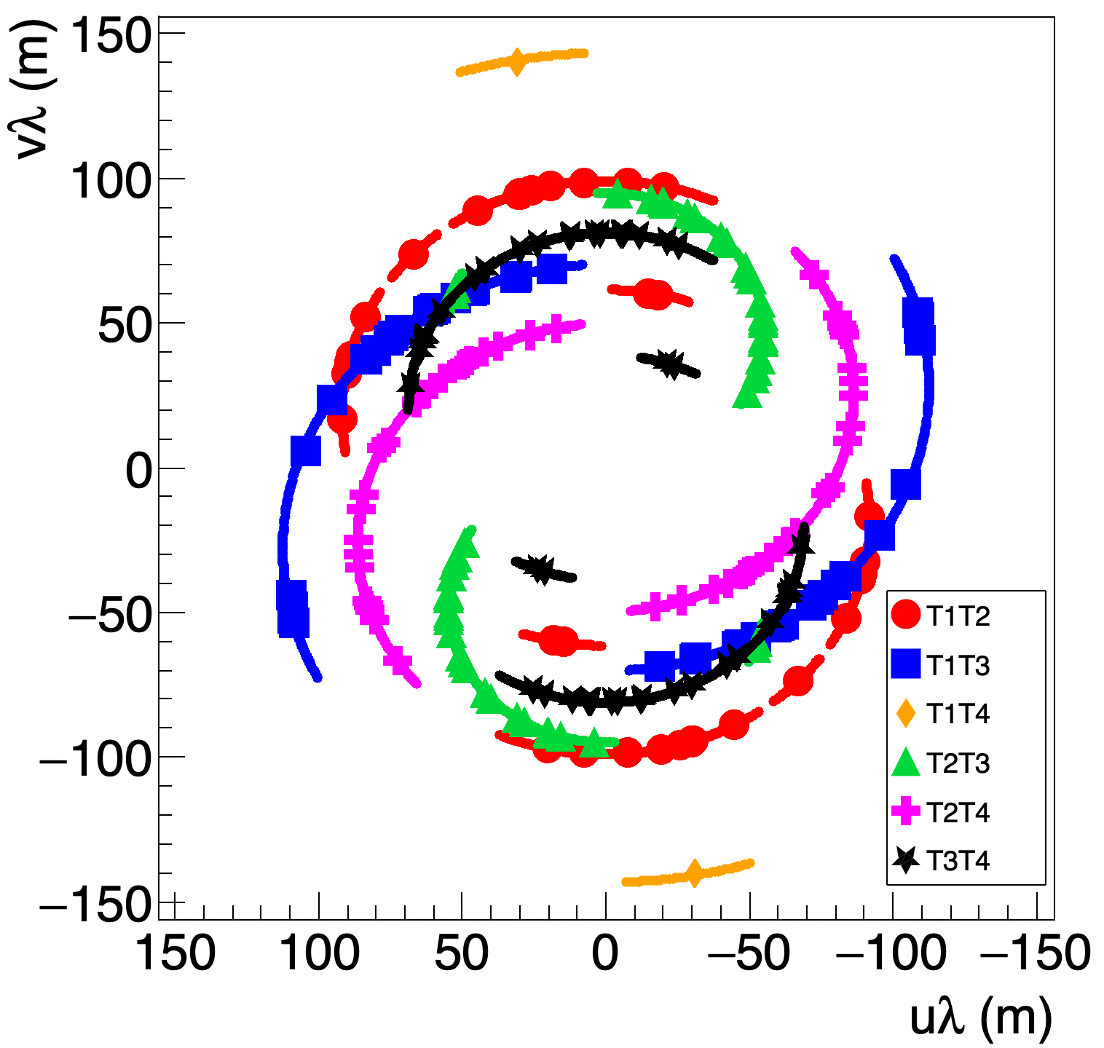}
    \caption{Baseline coverage of the six telescope pairs in the $u-v$ plane
    for the data tabulated in Table~\ref{tab:alldata}. 
    Points reflect the weighted-mean position of each $\sim$hour-long run.  The curves are point-reflected about the origin.}
    \label{fig:uvCoverage}
\end{figure}

VERITAS is composed of four 12-m diameter IACTs, usually
used for very high-energy ($E>100\,\rm GeV$) gamma-ray astronomy~\citep{HOLDER2006391}. 
During bright-moon periods, the system is used for SII measurements.
The four telescopes are individually referred to as T1, T2, T3, and T4, and are unequally spaced with separations ranging between about 80 and 170 meters, and interferometric projected baselines ranging between about 35 and 170 meters.
We have presented further details of the VSII stellar intensity 
interferometer in previous
publications~\citep{natureSIIDemo,VERITAS:2024quv}.
The optical light from the observed star is  detected by a single Hamamatsu super-bialkali R10560 photomultiplier tube (PMT)
located at the telescope focal plane, and continuously digitized at a 250 MHz rate with 8 bit resolution.
We use a narrow-band filter ($\lambda=416$) 
with a bandpass of 10 nm
to focus on the photosphere of the star.
An upgrade associated with the present study is the use of a
CPU-based server that simultaneously correlates all six
telescope pairs with a frequency-domain algorithm.  This
correlator produces correlations identical to the FPGA-based
correlator using a time-based algorithm that was used in
our previous measurements~\citep{ScottThesis:2023}.
For some observations in the present paper, a 10-MHz clock
improved the start-time
synchronization.

\begin{deluxetable*}{cccc}[t!]
\label{tab:GamCasbservations}
\tablecaption{VSII Observations of $\gamma$ Cassiopeiae}
\tablecolumns{4}
\tablewidth{0pt}
\tablehead{
\colhead{UTC Date} &
\colhead{Telescope Pairs Used} &
\colhead{Observation Time} &
\colhead{Target - Moon Angle}\\
\colhead{(yyyy-mm-dd)} &
\colhead{($T_A$,$T_B$)}& 
\colhead{(hours)} &
\colhead{(degrees)}}
\startdata
  2023-01-08 & (1,3), (2,3), (2,4), (3,4)  & 4.5 & 79 \\
  2023-02-02 & (1,2), (1,3), (2,3), (2,4), (3,4)  & 3.25 & 62 \\
  2023-12-25 & (2,4) & 2.0 & 51 \\
  2023-12-26 & (1,2), (1,3), (2,3), (2,4), (3,4)  & 7.28 & 56 \\
  2023-12-27 & (1,2), (1,3), (2,3), (2,4), (3,4)  & 6.0 & 63 \\
  2024-02-19 & (1,2), (1,3), (1,4), (2,3), (2,4), (3,4)  & 2.5 & 59 \\
  2024-02-20 & (1,2), (1,3), (2,3), (2,4)         & 2.0 & 66 \\
  2024-02-21 & (1,3), 2,3)         & 1.0 & 74 \\
  2024-02-22 & (1,3), (2,3), (2,4), (3,4)          & 1.75 & 82 \\
  2024-05-18 & (1,2), (2,3), (3,4)         & 1.0 & MBH \\
  2024-05-19 & (1,2), (2,3), (3,4)         & 1.0 & MBH \\
  2024-12-13 & (1,2), (1,3), (2,3), (2,4), (3,4) & 2.0 & 46 \\
  2024-12-15 & (1,2), (1,3), (2,3), (3,4) & 1.0 & 55 \\
  2024-12-17 & (1,3), (2,3), (2,4), (3,4) & 1.0 & 70 \\
\enddata
\tablecomments{MBH denotes moon below horizon. \\
\noindent
Dates and durations of VSII observations of $\gamma$~Cas are listed, together with the Moon angle and the telescope pairs used.
The four VERITAS telescopes are identified by numbers from one to four. The Moon angle is the angle between the Moon and the observed star. It correlates with the night sky background level affecting the data. Only a subset of telescope pairs from each observation were used in the analysis because some telescopes were not working, they had too large of a baseline, or the data was too noisy.} 
\end{deluxetable*}

Table~\ref{tab:GamCasbservations} lists the $\gamma$ Cas observations
for which data were collected, over 6 bright moon periods between January 2023 and December 2024.
Intensity interferometry is generally robust against ambient light for bright stars; VSII operations are reliable for target-to-moon angles from $30^\circ$ to $95^\circ$ \citep{VERITAS:2021xyw}.
Target-to-moon angles are listed in table~\ref{tab:GamCasbservations}.

Over 160 pair-hours of data were used, after quality cuts discussed below.
The baseline coverage in the star image Fourier reciprocal
($u-v$) plane is shown in figure~\ref{fig:uvCoverage}.
The relative telescope separation lies in the $u-v$ plane, the celestial east-north plane, following general interferometric convention. (See~\S~4 of~\citet{thompson2001} and \citet{DRAVINS2013331} for a further description of the $u-v$ plane.)

\section{Data Analysis}
\label{sec:Analysis}

\subsection{Overview}  

The size and shape of the photosphere are determined by measuring the dependence of the intensity correlation on the length and orientation of the projected baseline 
(i.e. separation between the telescopes perpendicular to the direction of the starlight) and then fitting this data to a model.
In this section, we discuss the steps taken to measure the intensity
correlation between pairs of telescopes; they are similar to those we have presented previously~\citep{natureSIIDemo,VERITAS:2024quv}.

The relative-time intensity correlation is measured as a function of
time during a run as described in section~\ref{sec:correlogram}.
After removing effects of
night-sky background light (section~\ref{sec:OffRuns}) 
radio interference, and tracking
corrections (section~\ref{sec:BadFrames}), the magnitude of the correlation
is quantified by a Gaussian fit to the intensity correlation
projected over a run, as discussed in section~\ref{sec:VisibilitiesAndUncertainties}.
Finally, in section~\ref{sec:Threshold} we identify some runs for which the noise is too large
to find a correlation peak.

\subsection{The correlogram}
\label{sec:correlogram}

The current $I(t)$ from the photomultiplier tube in each telescope 
is digitized and recorded at 250~MHz, i.e. in 4-ns samples.
A correlogram is formed by convoluting the digitized waveforms 
 from two telescopes $A$ and $B$ and normalizing by the product of the average values of the currents:

\begin{equation}
\label{eq:correlelogram}
    C_{AB}^{*}\left(\tau,T\right) \equiv
    \frac{\langle I_A(t)I_B(t-\tau-\tau_0) \rangle_t(T)}
    {\langle I_A(t)\rangle_t(T) \times \langle I_B(t)\rangle_t(T)} ,
\end{equation}
where $\langle\cdots\rangle_t$ indicates an average over time and
$\tau$ is the relative time (``time lag'') between samples.
The superscript ($^{*}$) denotes the correlation before correcting for effects
  of stray light, as discussed in section~\ref{sec:OffRuns}.
The constant $\tau_0$ is set equal to the optical path delay
between the telescopes at the beginning of the observation run.
The normalized correlation is constructed every 2~seconds of
time-in-run $T$.  We refer to these two-second intervals as ``frames'' below.

\begin{figure}[h!]
    \centering
    \includegraphics[width=0.5\linewidth]{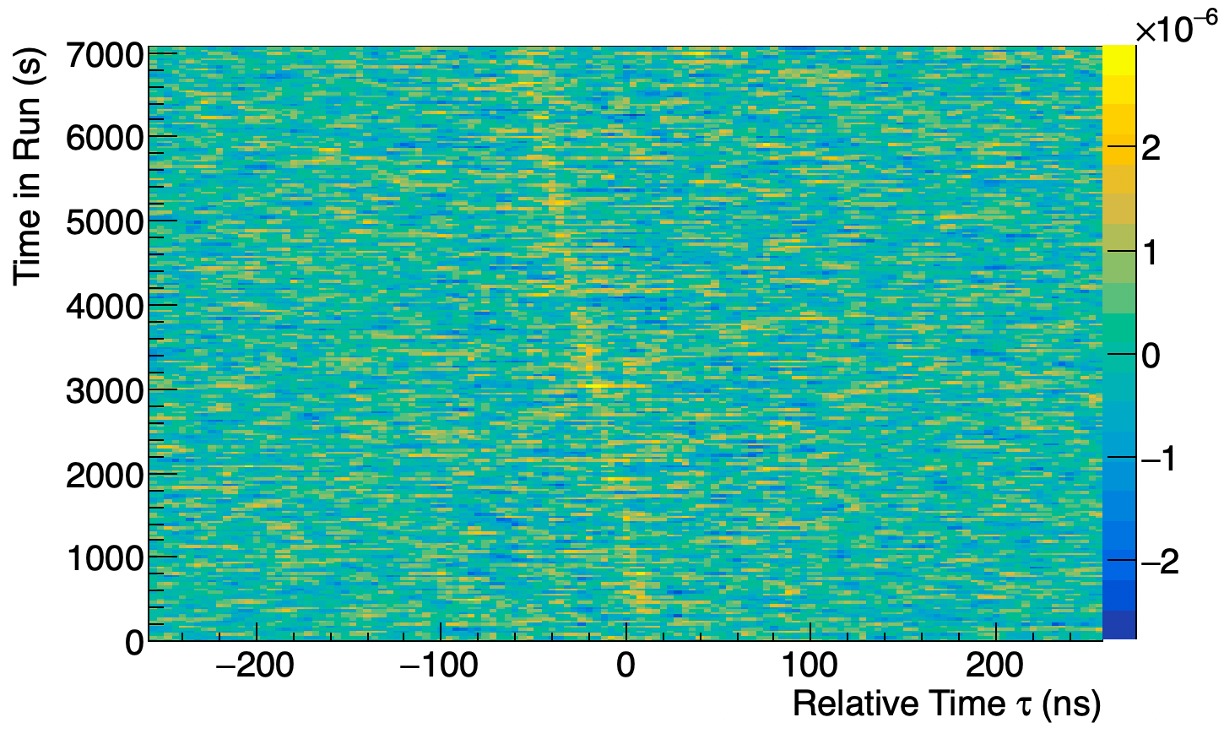}
    \caption{The correlation function as a function of relative time
    and time-in-run, for a typical 2 hour run.
    While the correlogram is constructed every 2~seconds, the vertical
    axis has been binned more coarsely in this figure, to reduce
    statistical noise and make the correlation more visible. 
    The diagonal feature is the correlation peak proportional to the squared visibility that changes during the run due to the variation in the optical path delay between the two telescopes.
    This correlation function is from pair T2T4, taken at 03:37 on UTC 2023-12-26. }
    \label{fig:HeatMap}
\end{figure}

A correlogram for a 2-hour observation run with telescopes T2 and T4 
is shown in figure~\ref{fig:HeatMap}.
A clear correlation signal is visible close to $\tau=0$ at the beginning of the run,
due to proper adjustment of $\tau_0$ (which we shall drop from further discussion for simplicity) in the correlator. 
The position of the correlation in relative time changes over the course of the run,
as the star moves across the sky, which changes the optical path delay between the fixed telescope-pair positions.

\subsection{Correction for residual light from other sources}
\label{sec:OffRuns}

During observation runs, the current in the phototubes is dominated by the light from
  the star.
Some of the light incident on the phototubes does not come from the star of interest, but
rather from the night sky background.
This uncorrelated contribution leads to a reduction in the raw measured correlation signal~\citep{10.1093/mnras/stt123}.
Neglecting this contribution would generate a run-by-run variation in the proportionality constant between the measured integral of the Gaussian fit of the correlation peaks and the actual squared visibility. 

We correct for this effect as we have done previously~\citep{natureSIIDemo,VERITAS:2024quv}.
Immediately before and after every observation run, the telescopes are pointed $0.5^\circ$ 
north of
the star and the average currents are recorded.
Depending on viewing conditions, stellar position, and moon angle and phase for a given observation run, the off-star current is $1-15\%$ of the on-star current.

The corrected correlogram $C(\tau,T)$ is related to the raw correlogram $C^{*}(\tau,T)$
 (equation~\ref{eq:correlelogram}) as
\begin{equation}
    \label{eq:DarkRunFactor}
    C_{AB}(\tau,T)-1 = \left(C^{*}_{AB}(\tau,t)-1\right)
    \cdot\frac{\langle I_A(t)\rangle_t(T)\,\langle I_B(t)\rangle_t(T)}
    {\left(\langle I_A(t)\rangle_t(T)-\langle I^{*}_A(t)\rangle_t(T)\right)\,
     \left(\langle I_B(t)\rangle_t(T)-\langle I^{*}_B(t)\rangle_t(T)\right)}, 
\end{equation}

where $\langle I^{*}_A(t)\rangle_t(T)$ represents time-averaged 
stray-light current at time-in-run $T$.
We assume that during the observations on the star itself, the ambient contribution
  varies linearly with time between what we measure before and after the on-star run.
Neglecting the stray light contamination would lead to a $\sim10\%$
  inaccuracy on the angular diameter estimate.

\subsection{Removal of artifacts from tracking corrections and
radio-frequency contamination}
\label{sec:BadFrames}

\begin{figure}[t]
    \centering
    \includegraphics[width=0.5\linewidth]{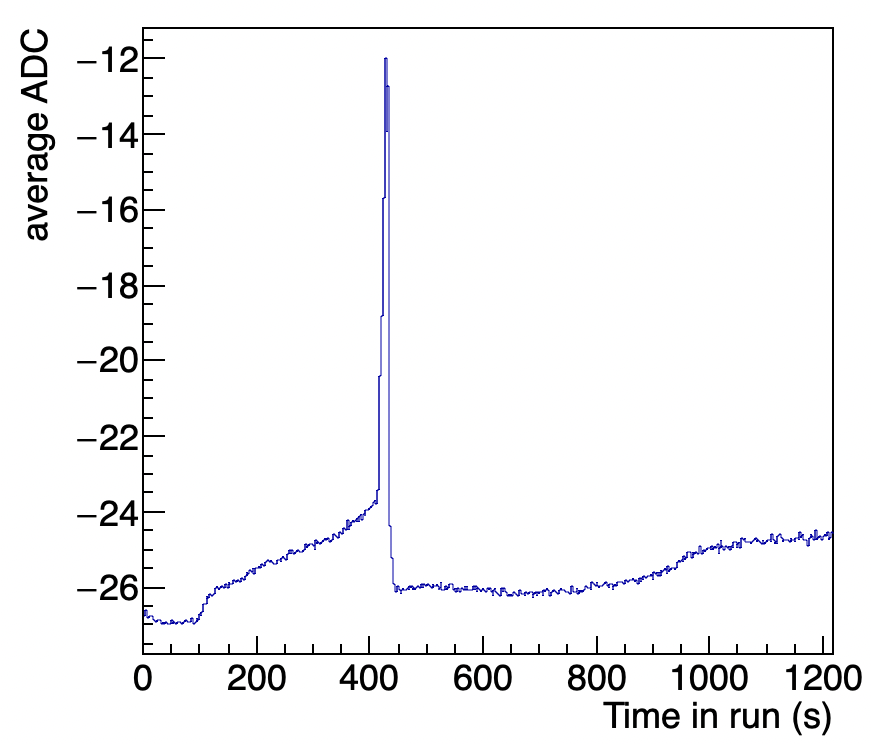}
    \caption{An example of a tracking correction in T2.  A large negative
    ADC 
    (analog-to-digital converter) 
    value corresponds to intense light incident on the PMT. 
    Due to tracking limitations of the array, at about 200~sec into the run, the
    star's image becomes slightly less centered on the 
    PMT.  
    At 400~sec
    into the run, the operator adjusted the tracking, causing the
    telescope to briefly stop tracking the star altogether, and the
    current drops sharply.  When the tracking is corrected, the current
    is stronger.  
    The frames around the spike are removed from the
    correlogram before processing further.
    \label{fig:BADTracking}
}
\end{figure}

\begin{figure}
    \centering
    \includegraphics[width=0.5\linewidth]{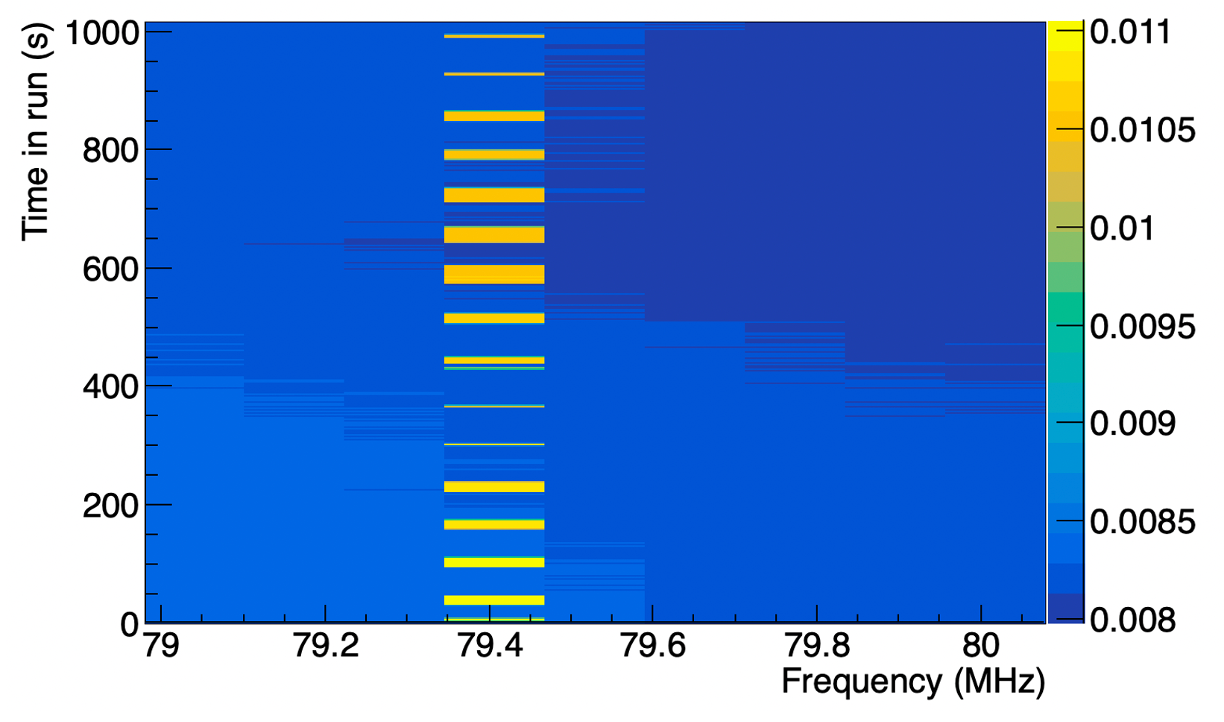}
    \caption{A portion of the power spectrum for T1, as a function
    of frequency and time in run.  Short, sporadic bursts from the on-site radio
    repeater produce strong signals at 79.4~MHz (an alias of the
    171~MHz radio signal).  Frames containing these bursts are
    removed from the correlogram before further processing.}
    \label{fig:PowerSpectrum}
\end{figure}

Some of the 
frames of the correlogram must be discarded
before further processing, due to two effects.

The first comes from telescope tracking corrections.  Occasionally
(typically once an hour), the image of the star moves slightly off
of the face of the photomultiplier tube.  When this happens, the
operator slightly adjusts the tracking parameters.  During this
procedure, the telescope briefly stops tracking altogether, and
the current drops sharply, as starlight is no longer being registered.
An example is seen in figure~\ref{fig:BADTracking}.

The second problem arises from an on-site radio repeater at  FLWO,
which transmits a 50~W signal at about 171~MHz in short ($\sim5$ sec), 
sporadic (roughly every 10-30~seconds) bursts.
This results in a strong aliased $79.4\,\rm MHz$ sinusoidal correlation-- 
more than 50 times
larger than the physical correlation of interest-- in all telescope
pairs~\citep{ScottThesis:2023}.
In addition to producing the correlogram, the correlator server
calculates the frequency power spectrum of each telescope for each
frame, and the 79.4~MHz spikes are easily identified, as demonstrated
in figure~\ref{fig:PowerSpectrum}.
Roughly one third of the frames are contaminated by this signal and
are removed from the correlogram.
Starting in late December, 2024, we received permission to disable
the repeater during VSII running, resulting in cleaner data and
effectively
about 40\% greater data collection efficiency.

Finally, for many runs and telescopes, there are much smaller frequency
spikes at 93.7, 99.6, and 107.5~MHz present in {\it every} frame.
The presence and strength of these signals depends on telescope orientation and vary from night to night.  They are likely due to
local radio stations.  When observed, the resulting sinusoidal
correlations are subtracted from the correlogram, frame-by-frame, 
as discussed in~\citet{VERITAS:2024quv}.

\begin{figure}[h!]
    \centering
    \includegraphics[width=0.5\linewidth]{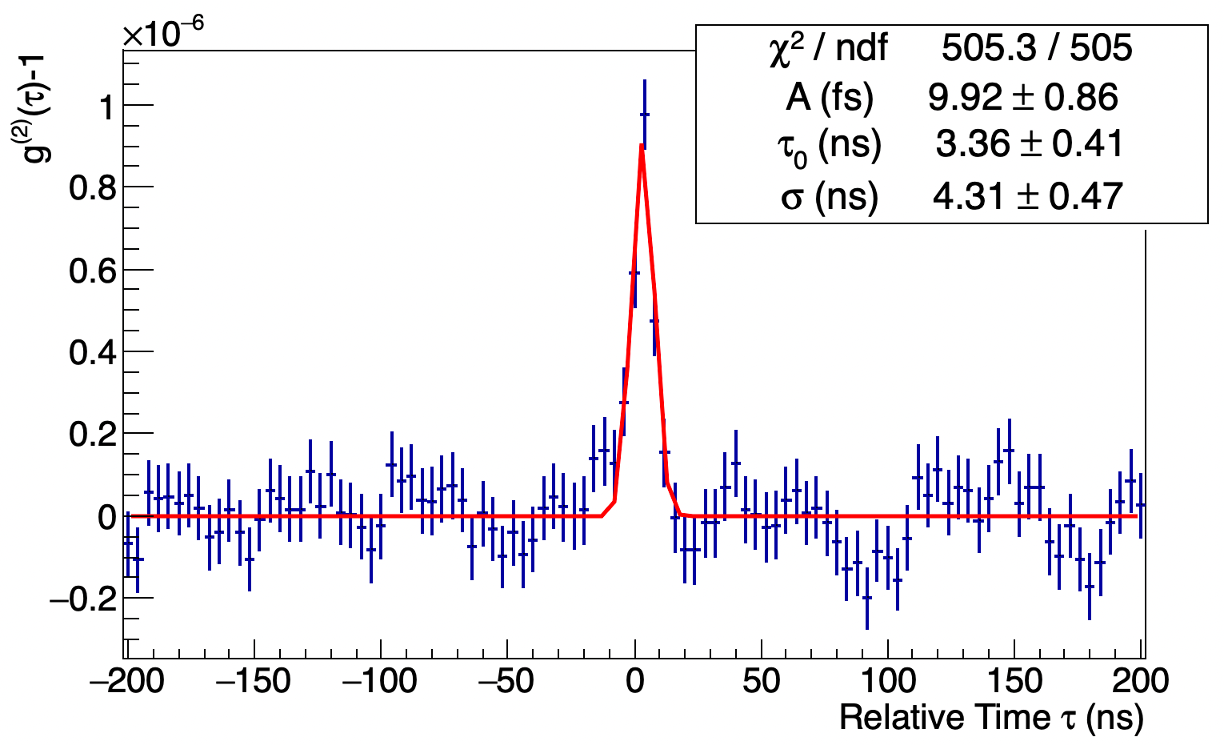}
    \caption{The OPD-corrected projection of the correlation function
    as a function of relative time, defined in equation~\ref{eq:g2}.  The correlation is quantified
    by the integral of a Gaussian peak fitted near relative time $\approx0$. This correlation function is from pair T2T3, taken at 02:31 on UTC 2024-02-19. 
    Before further analysis, the uncertainty on the area of the Gaussian fit ($A$) is replaced as discussed in section~\ref{sec:VisibilitiesAndUncertainties}.}
    \label{fig:HBTpeak}
\end{figure}

\subsection{Obtaining squared visibilities and their uncertainties}
\label{sec:VisibilitiesAndUncertainties}

Once the bad frames are removed and radio-station signals subtracted,
  a one-dimensional correlation function as a function of relative
  time is produced by shifting each frame by the appropriate optical
  path delay ($OPD$) and averaging over frames, according to
\begin{equation}
    \label{eq:g2}
    g^{(2)}(\tau) \equiv \langle C(\tau-OPD(T),T)\rangle_T .
\end{equation}
An example is shown in figure~\ref{fig:HBTpeak}.
A clear signal is observed close to relative time $\tau=0$.
Due to uncertainties on the clock cycle and exact start time, the
peak position may vary on order 10~ns, from one run to another~\citep{VERITAS:2024quv}.

A Gaussian fit is performed on $g^{(2)}(\tau)-1$ within a window $|\tau|<14$~ns. 
 An actual  correlation signal is expected to have a width around $4\,\rm ns$, due to the
finite sampling and response time of the photomultiplier tube and preamp.
If the peak-finder fails to find a peak in the window, or if its 
Gaussian width
is less than 2\,ns or greater than 6\,ns, then the data is discarded.
Variation of the $|\tau|$ window between $|\tau| < 10$ ns and $|\tau| < 20$ ns has a negligibly small effect on the measurements. The peak width cut was varied between 1.5 and 7 ns, and the effect of these variations is included in the systematic uncertainty estimate below.
The integral of the Gaussian, measured in units of time and
denoted $A$ in Figure~\ref{fig:HBTpeak}, is proportional to the squared visibility \citep{VERITAS:2024quv}.
The proportionality constant, determined by the bandwidth of
the optical filter, 
electronic response, and telescope optics, is expected~\citep{NolanMatthewsThesis}
to be on order $10^{-6}$~ns.  We treat this normalization constant as a parameter in our fits below.

The statistical uncertainty of this integral could be derived from the
fit parameters, but this ignores the residual correlations present in
the background.
To arrive at a better estimate of the uncertainty that accounts for
the residual background, we randomly add false
peaks, with the same area as the 
fitted
peak, on top of the background
far from the signal region ($|\tau|>50$~ns).
We then perform the same Gaussian fitting and selection procedure on
these false peaks as we do for the true peak.
The standard deviation of the distribution of false-peak Gaussian 
integrals is
taken as the uncertainty of the 
fitted peak near $\tau\approx 0$.
The uncertainty extracted in this way is typically about 40\% larger
than what would be obtained if we ignored the residual correlated background. 
This procedure is illustrated in more detail in Appendix \ref{sec:appendix errorbars}. 

We have reanalyzed the data of our previous measurement~\citep{VERITAS:2024quv} of the slow rotator
$\beta$~UMa, using our improved method of uncertainty estimations and removing tracking corrections.
The resulting angular diameter for a uniform-disk fit
changes by only 1\% (well within stated uncertainties),
but the $\chi^2$/ndf is reduced from 2.30 to 1.04.

\subsection{Exclusion of sub-threshold peaks}
\label{sec:Threshold}
In the presence of statistical and
residual correlated noise, any peak-finding algorithm finds spurious peaks when the actual signal is too small.
To identify the threshold at which this occurs, we used the peak-finder in random regions on the background far from the actual peak region ($|\tau|>50$~ns),
again applying the same Gaussian fitting and selection criteria as
for the actual signal peaks.

\begin{figure}[t!]
    \centering
    \includegraphics[width=0.6\linewidth]{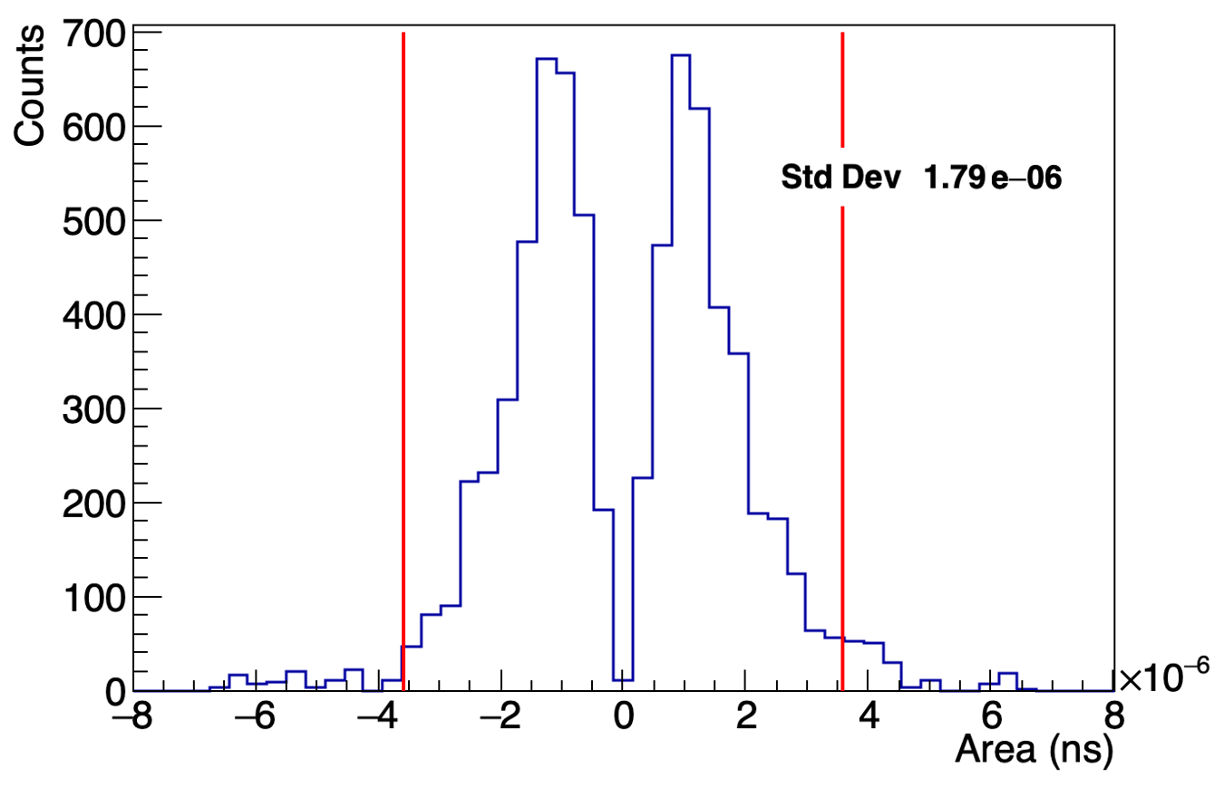}
    \caption{Distribution of areas from fits to background noise in the region of relative time far from the peak, 
    for all runs. The gray band on fig. \ref{fig:VisVersusBaseline} corresponds to the red lines at $\pm 3.58$~fs, which are $\pm2\sigma$ from the mean of this distribution. This measures the spread of areas that may be found when fitting to background noise alone, and includes 95\% of the data in the distribution. }
    \label{fig:branchingDist}
\end{figure}

The resulting distribution of Gaussian integrals is shown in 
figure~\ref{fig:branchingDist}.
As we have discussed previously~\citep{VERITAS:2024quv}, this
results in a bimodal distribution with a strong dip at zero.
We take twice the standard deviation of this distribution as a threshold, which includes 95\% of the fits \citep{RoseThesis2025}. 
Fitted peaks with integrals less than this threshold are indistinguishable
from noise and are discarded when extracting the size and shape
of the star, as discussed in sections~\ref{sec:SimpleGeometricModelFit}
and~\ref{sec:Modeling}.

\begin{figure}[t!]
    \centering
    \includegraphics[width=0.8\linewidth]{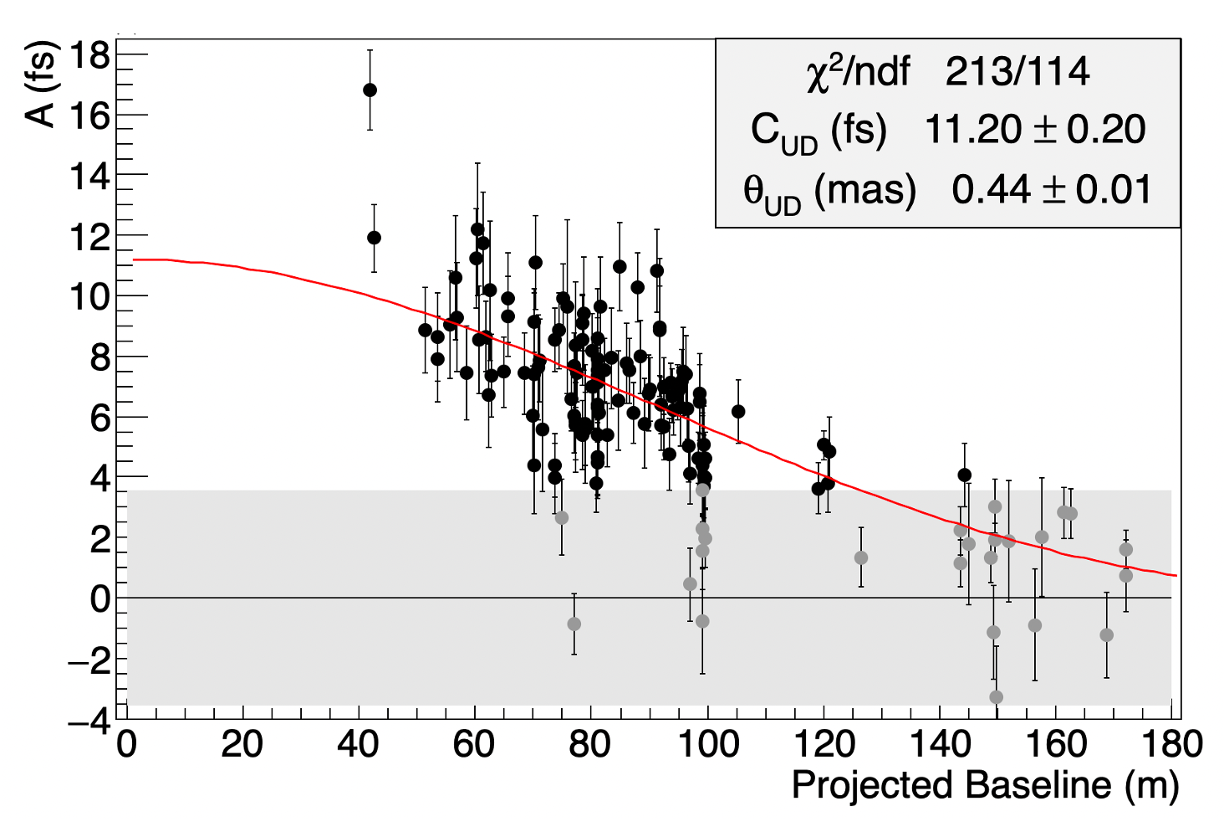}
    \caption{The area of the fitted correlation peak, proportional to
    the squared visibility, as a function of the magnitude of the
    baseline.  The curve represents the best fit with a uniform disk
    model. Points in gray, excluded from the fit, are below the threshold where the signal/noise is too low to reliably detect peaks, as discussed in section~\ref{sec:Threshold}. This threshold is marked with a gray shaded box. 
    }
\label{fig:VisVersusBaseline}
\end{figure}

Figure~\ref{fig:VisVersusBaseline} shows our measured visibilities as a function of
  the projected baseline.
The gray points represent those that pass the peak selections discussed in 
 section~\ref{sec:VisibilitiesAndUncertainties} but which fall below the noise
 threshold, shown as a gray band in the figure.
These come from runs in which the noise is too large, or the signal too small, to
  distinguish actual signal peaks from the noise.
These gray points are not used in further analysis.
We have varied the threshold between 3.0 and 4.1~fs to obtain a systematic uncertainty on fit parameters discussed in section~\ref{sec:SimpleGeometricModelFit}.

\newpage

\section{Extracting the size and shape of the photosphere}
\label{sec:SimpleGeometricModelFit}

In this section, we fit the visibilities with simple models, to extract an effective size, shape, and orientation of the star.
It is common to model a star as a uniformly illuminated disk of angular radius $\theta_{UD}$.
In this case, the squared visibility for a projected baseline $b$ is 
\begin{equation}
\label{eq:UniformDiskModel}
    A(b) = C_{\rm UD}\times\left(\frac{2J_1(\pi b\,\theta_{\rm UD}/\lambda)}{\pi b\, \theta_{\rm UD}/\lambda}\right)^2 ,
\end{equation}
where $\lambda=416$~nm and $J_1$ is the first-order Bessel function. 
$C_{\rm UD}$ is the proportionality constant between actual squared visibilities and peak integrals measured from correlation functions, which we treat as a fit parameter.
The red curve in figure~\ref{fig:VisVersusBaseline} is the best fit of 
  equation~\ref{eq:UniformDiskModel} to the data.
The effective uniform disk angular radius is $0.44\pm0.01{\rm (stat)}\pm 0.02{\rm (syst)}$~mas,
consistent with the result reported by 
the MAGIC collaboration~\citep{10.1093/mnras/stae697} 
of $0.515\pm0.038{\rm (stat)}\pm0.023{\rm (syst)}$. Our systematic uncertainties were calculated by varying the peak quality cuts and threshold and not removing tracking corrections or accounting for the background light contributions. 
As discussed below, this star is not expected to be circular, so some tension between the values
  of $\theta_{UD}$ measured by different observatories may be expected. 

A rapid rotator like $\gamma$ Cas is expected have an equatorial bulge and anisotropic
  gravitational darkening, which would lead to an oblate source with
  spatial extension perpendicular to the axis of rotation.
Therefore, fitting the data with a uniform circular disk may not be appropriate, as the length scale may depend on the orientation of the baseline.
An intuitive way to explore this possibility is to measure the uniform disk effective angular size
  as a function of angle of the projected baseline in the $u-v$ plane, $\phi_b$.
The angle with the smallest effective angular diameter would correspond to the position angle of
  the star rotation axis.

\begin{figure}[t!]
    \centering
    \includegraphics[width=0.99\linewidth]{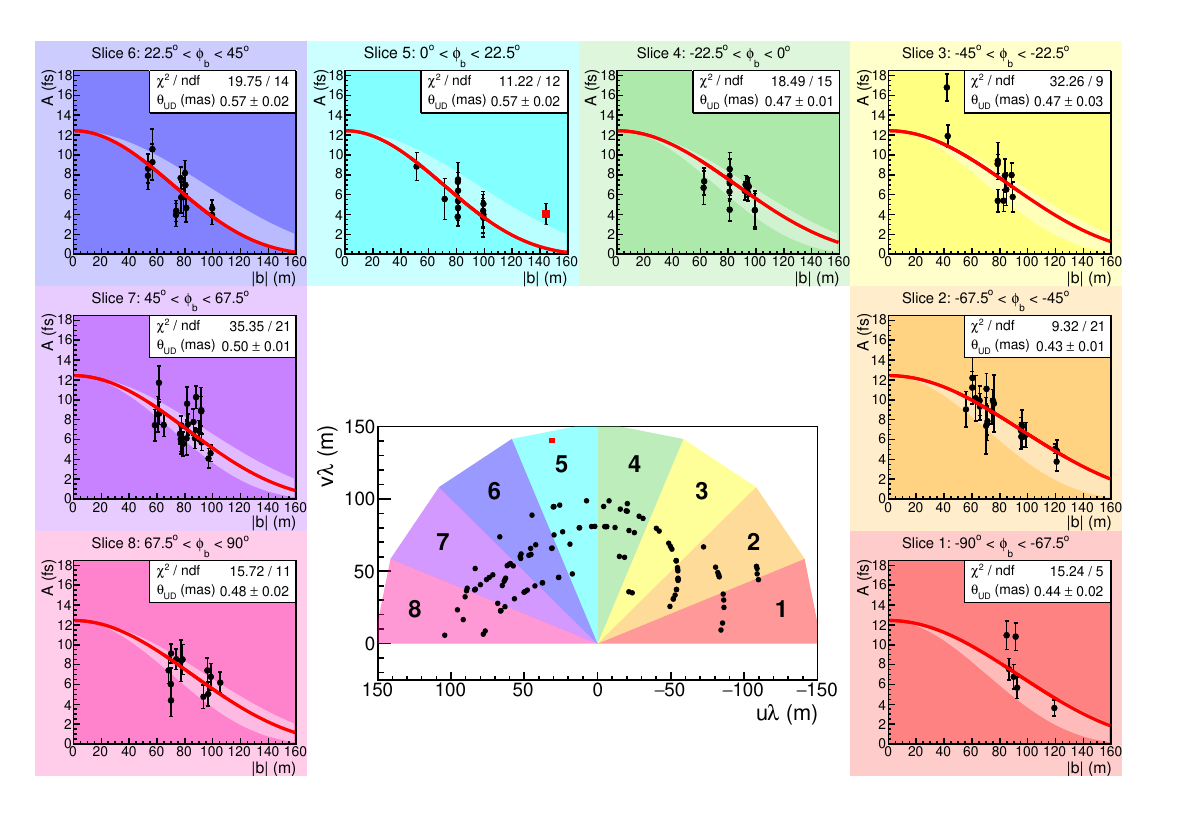}
    \caption{The center panel shows the $u-v$ midpoint of all runs and telescope pairs, 
  with different color shaded regions indicating angular slices in the $u-v$ plane. For data within each of the individual slices, the surrounding panels of corresponding colors show  the squared visibility as a function of the length of the baseline. Each slice is fit with a uniform disk model, to extract the effective 1-dimensional radius of a given angular projection.  
    The zero-baseline squared visibility is constrained to be the same for all slices, with a value from the uniform ellipse fit of figure~\ref{fig:UniformEllipseFit}. 
    Identical light shaded regions in each panel show the range of fits 
    between the most extreme values of the 8 individual slices; 
    these are drawn to guide the eye, making small differences between panels more apparent.
    A datapoint at large baseline in slice 5 is drawn in red and discussed in the text.
    }
    \label{fig:Pizza}
\end{figure}

\begin{figure}[t!]
    \centering
    \includegraphics[width=0.95\linewidth]{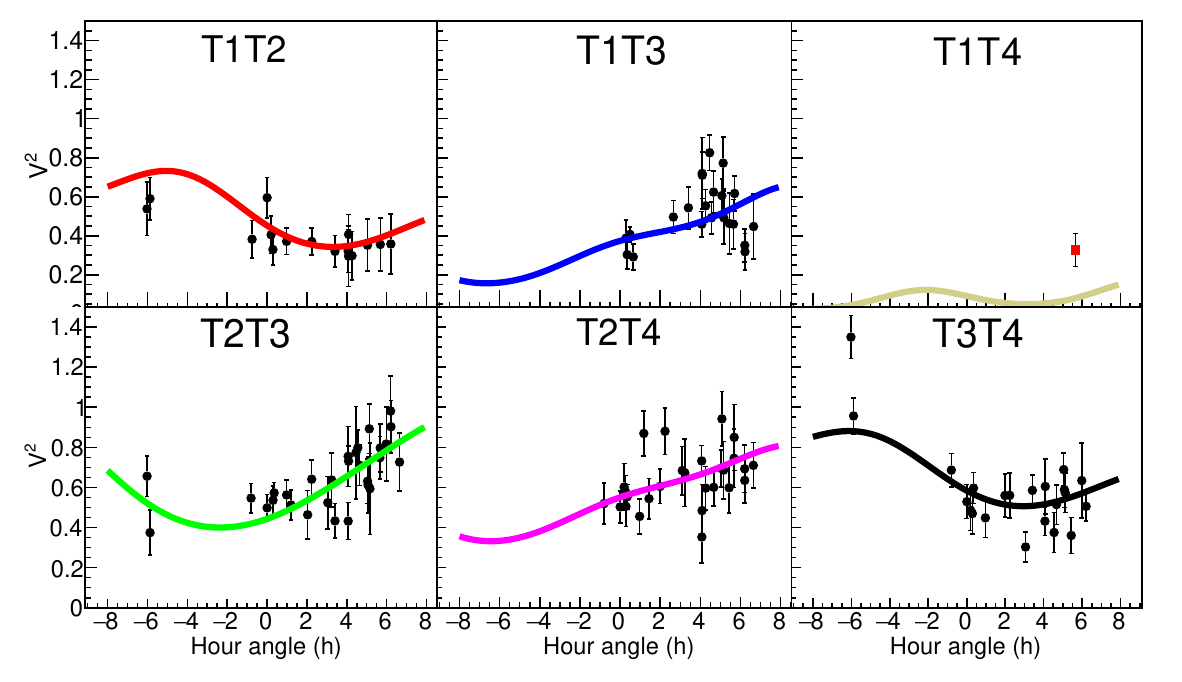}
    \caption{For each telescope pair, the normalized visibility is shown as a function of local hour angle.  Curves represent the best fit with a uniform ellipse model. The singular T1T4 data point is shown as a red square, because it passes all quality cuts yet we suspect it may be a fit to random background fluctuations, like those discussed in section \ref{sec:Threshold}. It is included in the fit shown.
    Fits are reported with and without this point in table~\ref{tab:analytic fit results}.}
    \label{fig:UniformEllipseFit}
\end{figure}

In figure~\ref{fig:Pizza}, the visibility measurements are sorted according to ranges in $\phi_b$.
In the center panel, the baseline for each measurement is shown in the $u-v$ plane, and $22.5^\circ$-wide slices are defined and shown as differently colored regions.
Measurements that fall within each slice are shown in the surrounding panels and fitted with the uniform disk model of equation~\ref{eq:UniformDiskModel} to extract an effective one-dimensional angular diameter for a small range in $\phi_b$.
While the visibility measurement from T1T4 with baseline of 145\,m passes our threshold cut (see figures~\ref{fig:VisVersusBaseline} and~\ref{fig:UniformEllipseFit}), we suspect it may be a fluctuation, like those discussed in section~\ref{sec:Threshold}.
Therefore, we report fit parameters obtained both including and excluding this point, and it is drawn as a red square in Figures \ref{fig:Pizza}, \ref{fig:UniformEllipseFit}, and \ref{fig:StellarModelFit}.

\begin{figure}[t!]
    \centering
    \includegraphics[width=0.5\linewidth]{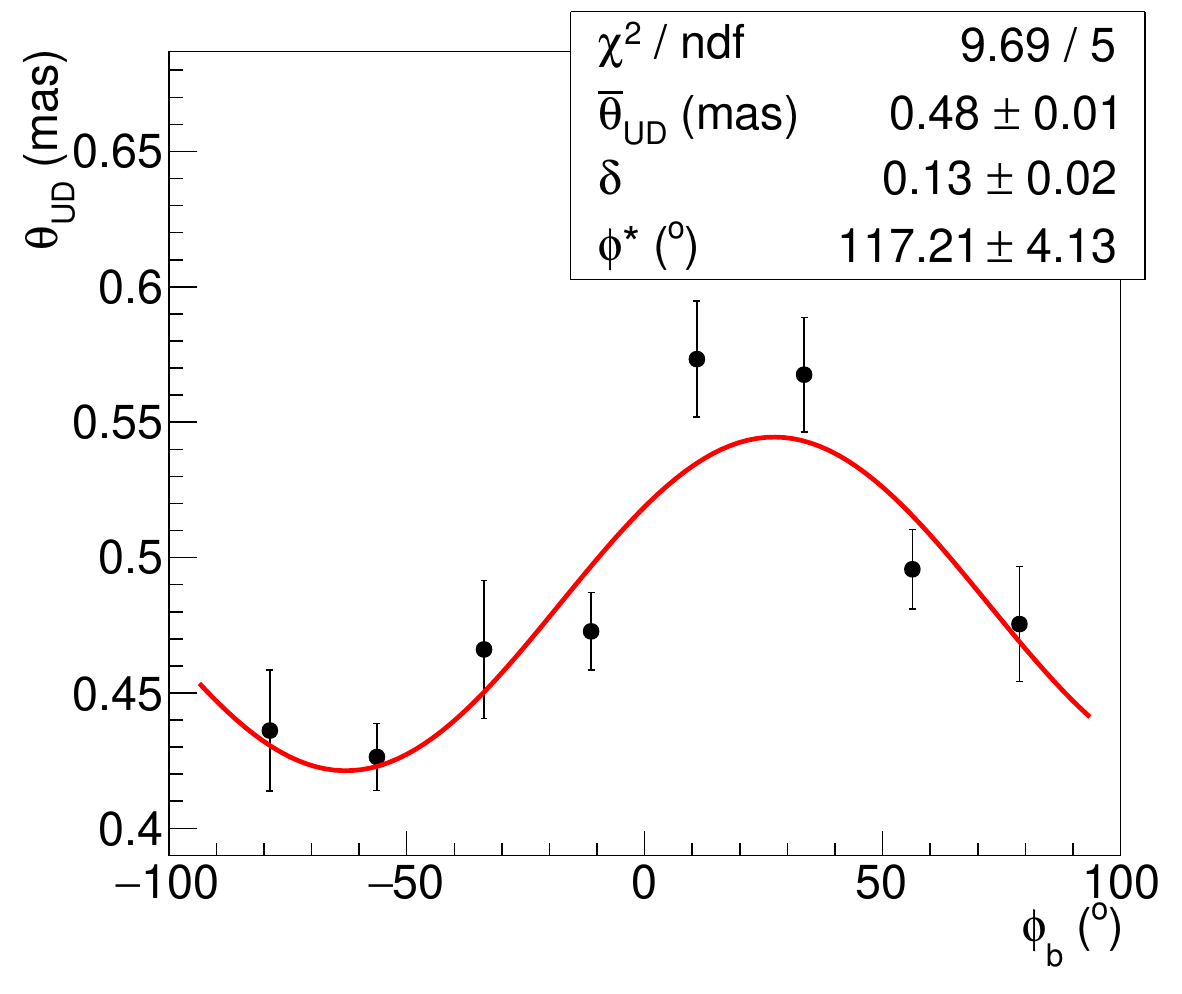}
    \caption{The uniform-disk radius from fitting $A$ versus baseline to datapoints in different slices in the $u-v$ plane (see figure~\ref{fig:Pizza}), as a function of the average angle of the slice.}
    \label{fig:PizzaRadii}
\end{figure}

Figure~\ref{fig:PizzaRadii} shows the effective angular diameter versus the average position angle
of the slice.
The data are well fit with the function
\begin{equation}
    \label{eq:PizzaRadii}
    \theta_{UD}(\phi_b) = \overline{\theta}_{UD} (1-\delta\cos(2(\phi_b-\phi^*))) .
\end{equation}
The average effective diameter is $\overline{\theta}_{UD}=0.48 \pm 0.01$~mas. 
The minimum diameter, the minor axis, is at $\phi_b=\phi^*=117^\circ\pm4^\circ$, which we associate with the  position angle
of the stellar rotation axis
and is consistent with previous measurements of the accretion disk of this star~\citep{sigut_2020}.
The extracted magnitude of the modulation, $\delta=0.13 \pm 0.02$ 
corresponds\footnote{If an oscillation $A\sin(2\phi)$ is uniformly integrated
over bins of width $\Delta$, the resulting histogram will exhibit
oscillations of magnitude $A\tfrac{\sin\Delta}{\Delta}$.  In our case, this is small effect:
the finite-binning-corrected oscillation magnitude is $0.13\tfrac{\pi/8}{\sin(\pi/8)}=0.133$.} 
to a maximum-to-minimum ratio of 
$\tfrac{1+\delta}{1-\delta}=1.30 \pm 0.05$.

While figures~\ref{fig:Pizza} and~\ref{fig:PizzaRadii} provide an intuitive way to 
estimate the anisotropy and orientation of the rapid rotator, it is more accurate
to perform a simultaneous fit of all visibilities as a function of $u$ and $v$. The hour angle, $H$, for a given source and observatory, uniquely defines the $u-v$ coordinates for each telescope pair, so we fit the data as a function of hour angle \citep{DRAVINS2013331}.
The data is fitted with a uniformly illuminated ellipse model~\citep{Tycner:2006nw}
\begin{equation}
    \label{eq:UniformEllipseModel}
    A(H) = A(u(H), v(H)) = C_{\rm norm} |V|^2 = C_{\rm norm}\left(\frac{2J_1\left(\pi \theta_{\rm min} s\right)}{\pi\theta_{\rm min}s}\right)^2 ,
\end{equation}
where
\begin{equation}
    s\equiv\sqrt{
    r^2(u\cos\phi^* - v\sin\phi^*)^2
    + (u\sin\phi^* + v\cos\phi^*)^2} .
\end{equation}
Here, $r$ is the ratio of the major to minor axes ($r\geq1$),
  $\theta_{\rm min}$ is angular diameter of the minor axis, 
  and $\phi^*$ is the position angle of the minor axis of the
  ellipse.\footnote{Note that~\citet{Tycner:2006nw} describes the model for $r\leq1$.}
For a rapid rotator, then, $\phi^*$ is the position angle of the axis of rotation.

The proportionality between the actual squared visibility and the peak integrals is $C_{\rm norm}$,
  which we treat as a fit parameter.
Figure~\ref{fig:UniformEllipseFit} shows the peak integrals (divided by $C_{\rm norm}$) 
for each telescope pair, as a function of local hour angle.  

This more complete model describes the data better than the uniform disk model discussed above; see table~\ref{tab:analytic fit results}.
The angular size of the minor axis is $\theta_{\rm min}=0.43\pm0.02{\rm (stat)}\pm 0.02{\rm (syst)}$~mas, and
the axis ratio is $r=1.28\pm0.04\pm0.02$.
Fitting an elliptical source with a uniform circular model (equation~\ref{eq:UniformDiskModel})
  would result in a uniform disk radius ($\theta_D$) somewhere between 0.43~mas and 0.55~mas, depending on the orientation of the baseline.
Therefore, the value of $\theta_{UD}$ extracted from any given measurement will depend
  on the orientation of the baselines used during observation.
This complicates comparisons between, say VSII and MAGIC~\citep{10.1093/mnras/stae697},
when using a circular source model on a rapid rotator. 

It also means that $\overline{\theta}_{UD}$ from equation~\ref{eq:PizzaRadii} and
  $\theta_{UD}$ from equation~\ref{eq:UniformDiskModel} represent slightly different
  quantities.
While figures~\ref{fig:Pizza} and~\ref{fig:PizzaRadii} are an intuitive way to display
  the non-circular nature of the star, the characteristics of the elliptical appearance of the star are most
  accurately extracted by a fit to an elliptical image model.

The extracted position angle of the rapid rotator star axis is estimated to be
  $\phi^*=116^\circ\pm5^\circ{\rm (stat)}\pm7^\circ{\rm (syst)}$, consistent with measurements of the decretion disk
  at longer wavelengths~\citep{sigut_2020}, 
  indicating that the decretion disc is generally aligned along the equatorial bulge of the photosphere itself.

\begin{deluxetable*}{ccccccc}[h]
\label{tab:analytic fit results}
\tablecaption{Parameters of simple models fits.} 
\tablecolumns{7}
\tablewidth{0pt}
\tablehead{
\colhead{T1T4 point} &
\colhead{Model} &
\colhead{$\phi^*$ (º)} &
\colhead{$\theta_{\rm min}$ (mas)} &
\colhead{$\theta_{\rm maj}/\theta_{\rm min}$} &
\colhead{$C_{norm}$ (fs)} &
\colhead{$\chi^2/ndf$}}
\startdata
 &
Uniform Disk & - & 0.44 $\pm$ 0.01 $\pm$ 0.03 & (1) & 11.2 $\pm$ 0.2 $\pm$ 0.5 & 213/114 \\
Excluded & uv slices & 117 $\pm$ 4 $\pm$ 6 & 0.42 $\pm$ 0.02 $\pm$ 0.02 & 1.30 $\pm$ 0.05 $\pm$ 0.03 & 12.4* & 9.69/5\\
 & Uniform Ellipse & 116 $\pm$ 5 $\pm$ 7 & 0.43 $\pm$ 0.02 $\pm$ 0.02 & 1.28 $\pm$ 0.04 $\pm$ 0.02 & 12.4 $\pm$ 0.5 $\pm$ 0.5 & 170/112\\
\hline
& Uniform Disk & - & 0.43 $\pm$ 0.01 $\pm$ 0.02 & (1) & 11.0 $\pm$ 0.2 $\pm$ 0.5 & 215/115 \\
Included & uv slices & 118 $\pm$ 4 $\pm$ 6 & 0.41 $\pm$ 0.01 $\pm$ 0.02 & 1.27 $\pm$ 0.05 $\pm$ 0.04 & 12.2* & 7.39/5\\
 & Uniform Ellipse & 117 $\pm$ 5 $\pm$ 8 & 0.42 $\pm$ 0.02 $\pm$ 0.02 & 1.27 $\pm$ 0.05 $\pm$ 0.03 & 12.2 $\pm$ 0.5 $\pm$ 0.5 & 178/113 \\ 
\enddata
\tablecomments{note that $\chi^2/ndf$ for the uv slices analysis is based on sinusoidal fit to round model fits of the 8 ``pizza slice" sections. \\
\noindent
*$C_{norm}$ for the uv slice model is fixed to $C_{norm}$ from the Uniform Ellipse model fit. \\
\noindent
Errorbars are reported as $\pm (stat) \pm (syst)$. }
\end{deluxetable*}

\section{Constraining a stellar model of a rapid rotator}
\label{sec:Modeling}

We have computed synthetic squared visibilities, high-resolution
spectra and spectral energy distributions for the photosphere of
$\gamma$ Cas using PHOENIX \citep[version 20.01.02B]{H99}, a code for
nova, supernova, stellar and planetary model atmospheres.  We use a
Roche–von Zeipel model for a rapidly rotating star
\citep{vega06,sack23}, modeled as an infinitely
concentrated central mass under uniform angular rotation.

Each axially symmetric model is parameterized by a trigonometric
parallax, $\varpi$, equatorial angular diameter, $\theta_{\rm eq}$,
polar effective temperature, $T_{\rm pole}$, polar gravity, $\log(g)_{\rm pole}$,
fraction of the critical (breakup) angular rotation rate, $\Omega/\Omega_{\rm c}$, and
the von Zeipel gravity darkening parameter, $\beta$.  As seen from
Earth, the model is further parameterized by the inclination $i$, and the position angle, $\phi^*$,
of the visible stellar rotation axis.

The intensity of each point on the surface of the model star, as
viewed from Earth, is interpolated from a grid of 106 PHOENIX stellar
atmosphere radiation fields: specific intensities at 78 direction cosines between surface normal and the direction of an observer and 106 ($T_{\rm eff}, \log(g)$) pairs
(with step sizes of 500 K in $T_{\rm eff}$ and 0.1 in $\log(g)$) 
between the hot, high gravity pole and the cooler, lower gravity equator. A model rotating very close to the critical limit
($\Omega/\Omega_{\rm c}$  = 0.9999) has a surface temperature difference of $\Delta T_{\rm eff}\simeq$ 12000 K and a gravity difference of $\Delta\log(g) \simeq 1.2 $.

For our model of $\gamma$ Cas we fixed $\varpi = 5.94$ mas \citep[$5.94\pm 0.12$ mas]{V07},
 $i = 60^\circ$ as constrained
by \citet{laily_sigut_2024} (see also
\citet{sigut_2020} and \citet{Tycner:2006nw}), and $\beta =
0.20$, as empirically determined for rapid rotating stars imaged with CHARA/MIRC \citep{che2011}.  

We computed 7750 models for $\gamma$ Cas with 31 $\theta_{\rm eq}$
values from 0.40 mas to 0.70 mas, 25 $\phi^*$ values from 90$^\circ$ to
138$^\circ$ and 10 $\Omega/\Omega_{\rm c}$ values from 0.9000 to
0.9999.  With six baselines and 118 hour angles, this corresponds to
a table of 5,487,000 synthetic visibility values.

\begin{figure}[t!]
    \centering
    \includegraphics[width=0.8\linewidth]{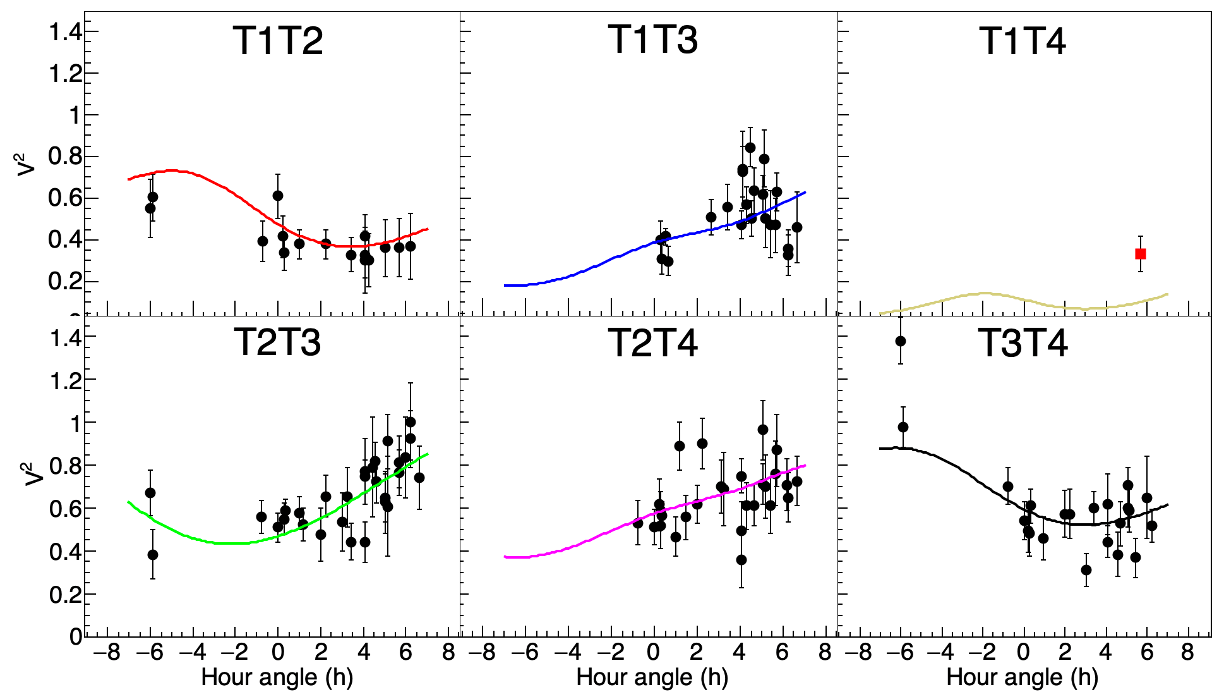}
    \caption{Using the same data shown in figure~\ref{fig:UniformEllipseFit}, the best fit using a
    Roche-von Zepiel models for a rapidly rotating star.} 
    \label{fig:StellarModelFit}
\end{figure}

\begin{figure}[p]
    \centering
    \includegraphics[width=1.0\linewidth]{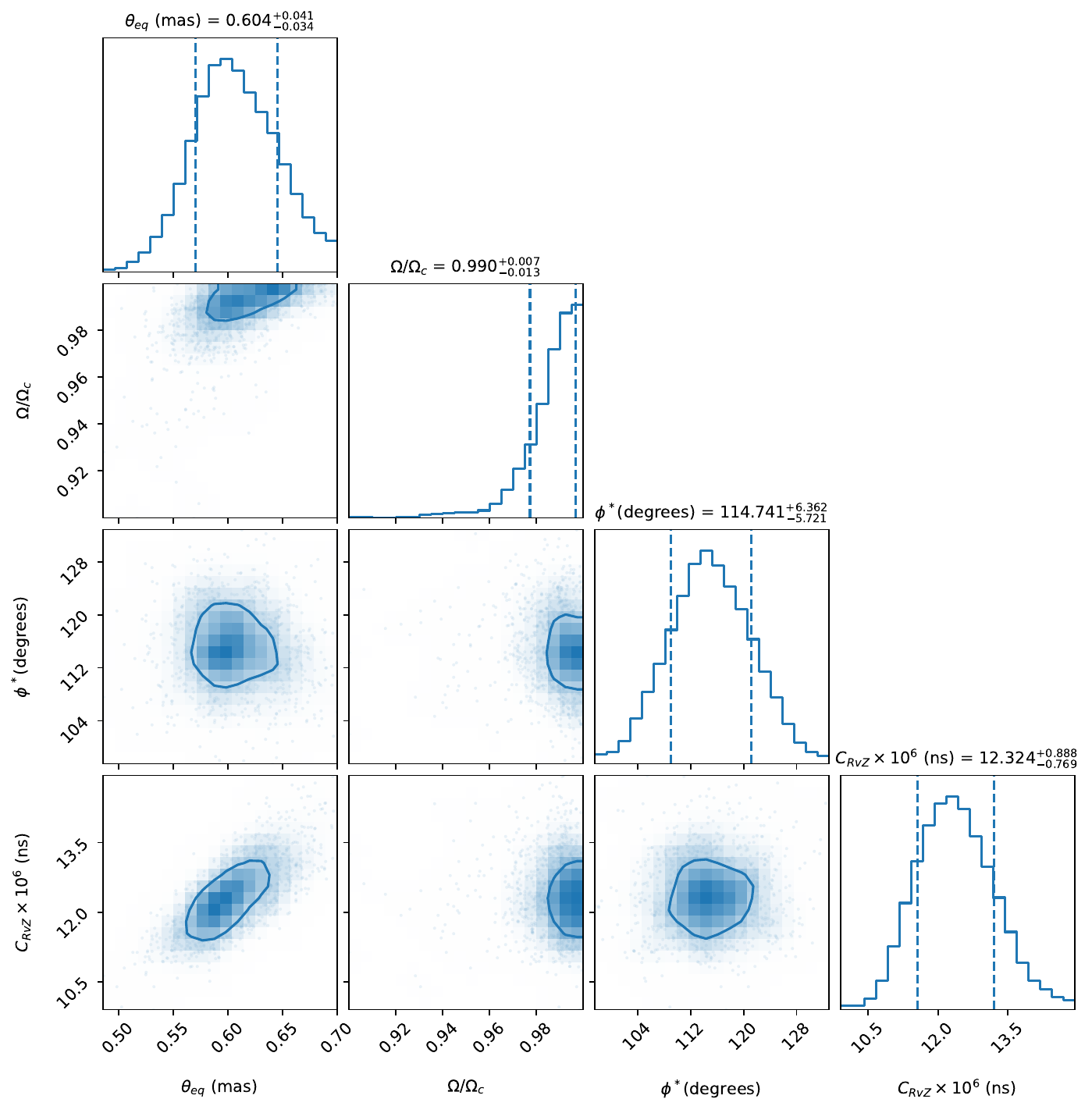}
    \caption{Best fit parameter distributions  for the Roche-von Zeipel stellar model from fitting 1000 bootstrap samples with replacement from Table \ref{tab:alldata}.  The dashed lines show the 1$\sigma$ lower bound error  and the 1$\sigma$ upper bound of each parameter: $\theta_{\rm eq}$, the equatorial angular diameter; $\Omega/\Omega_{\rm c}$, fraction of the critical angular rotation rate; $\phi^*$ position angle of the visible rotation axis; and $C_{RvZ}$, the proportionality constant between the model squared visibilities and measured correlation peak integrals. }
    \label{fig:RvZ_cornerplot}
\end{figure}

Figure~\ref{fig:StellarModelFit} shows the VERITAS data (the same as in Figure~\ref{fig:UniformEllipseFit}) with a Roche-von Zeipel stellar model.  Figure~\ref{fig:RvZ_cornerplot} shows the distributions of the best-fit parameters.  The $\chi^2/ndf$ fit to the data confronts the upper physical bound of unity, therefore we can only define a $1 \sigma$  lower limit on the rotation rate at 97.7\%

\begin{figure}[t!]
    \centering
\includegraphics[width=0.6 \linewidth]{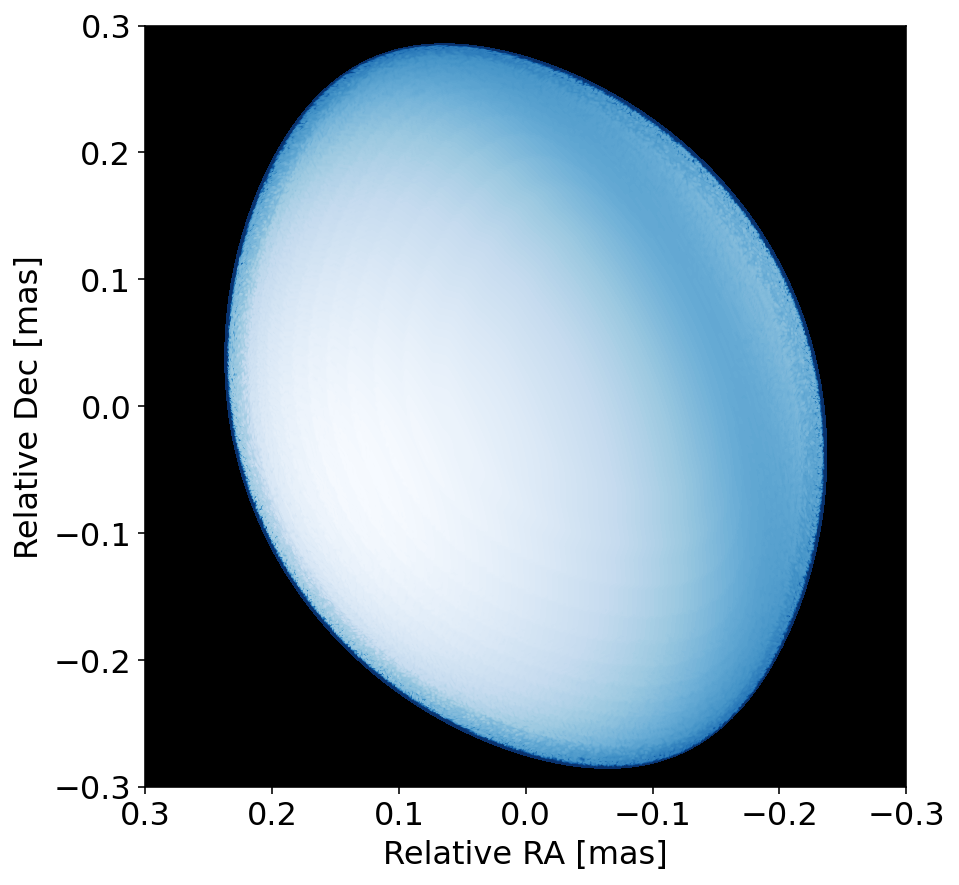 }
    \caption{A synthetic photosphere for $\gamma$ Cas with $\theta_{\rm eq}$=0.60 mas, $\Omega/\Omega_{\rm crit}$ =0.9888
     and $\phi^* = 114^\circ$ consistent with the best fit values to the VERITAS interferometry.} 
    \label{fig:Football}
\end{figure}

Figure~\ref{fig:Football} is a rendering of the calculated light intensity distribution corresponding to
the Roche-von Zeipel model that fits the data best.  

\begin{figure}[t!]
    \centering
    \includegraphics[width=1.0\linewidth]{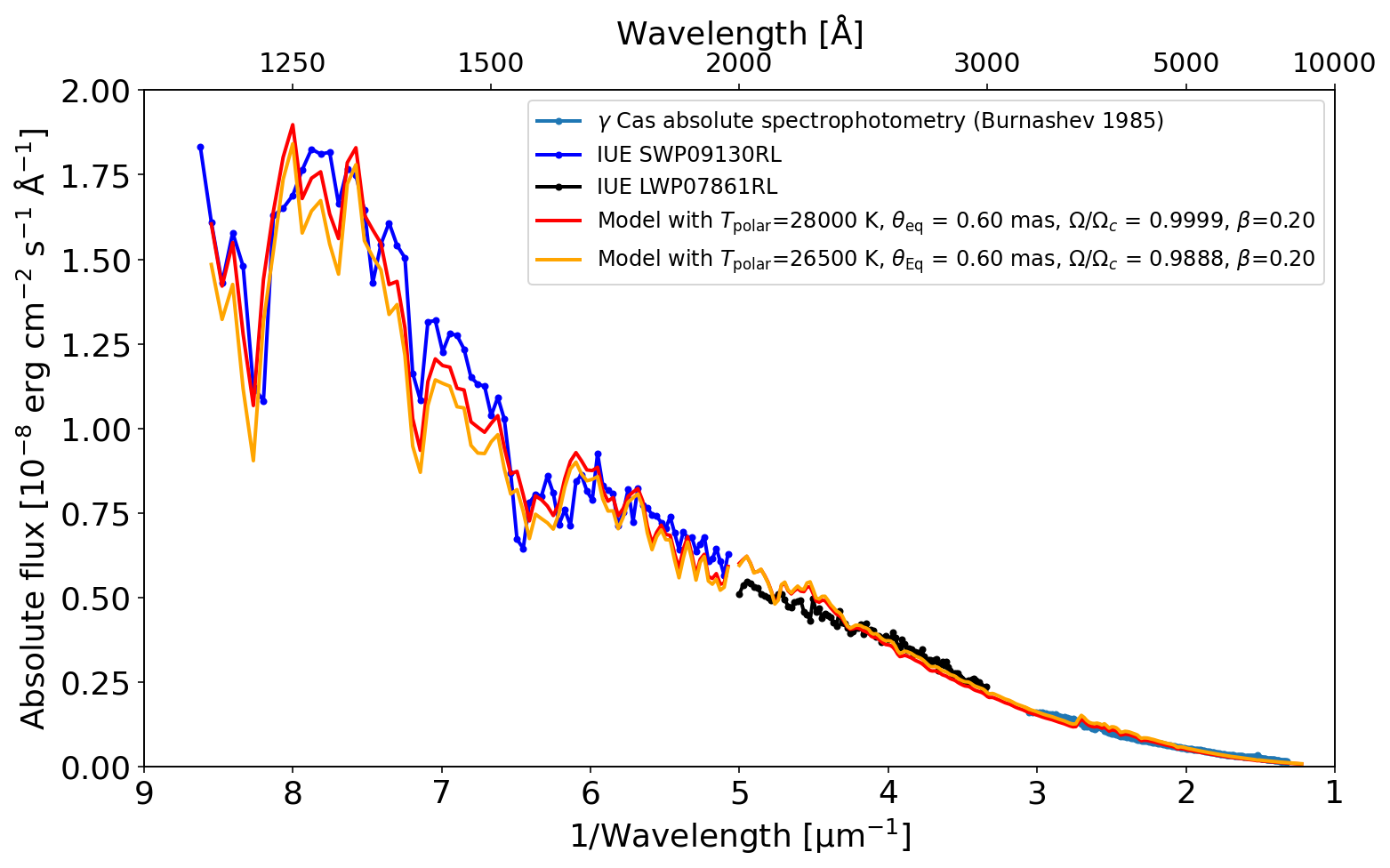}
    \caption{A comparison of two Roche-von Zeipel synthetic spectral energy distributions (SEDs) to archival absolute spectrophotometry of $\gamma$ Cas between 1200\,\AA\ and 8170\,\AA. Data for wavelengths beyond 3200\,\AA\ are from \dataset[Burnashev (1985)] {https://cdsarc.cds.unistra.fr/viz-bin/cat/III/126}. International Ultraviolet Explorer (IUE) data have been rebinned from high dispersion to low dispersion \citep{INES_IUE_data} and extracted from the \dataset[IUE Newly-Extracted Spectra (INES) data archive]{http://sdc.cab.inta-csic.es/ines/index2.html}. The faster spinning model ($\Omega/\Omega_{c}$ = 0.9999) is more gravity darkened and requires a larger polar temperature at the same equatorial angular diameter to match the spectrophotometry.} 
    \label{fig:SED_comparison}
\end{figure}

For the best-fitting median values,  $\theta_{\rm eq}$ = 0.60 mas and $\Omega/\Omega_{c}$ = 0.99,
$T_{\rm pole}$ = 26500 K provides a reasonable match to available absolute archival spectrophotometry in the UV optical range (see Figure~\ref{fig:SED_comparison}). A $\log(g)_{\rm pole}$ = 3.82 yields a mass of 15.1 M$_\odot$, at
the top of the range 13-15 M$_\odot$ adopted by \citet{myron2019}.
Table~\ref{tab:stellar_parameters} shows the best fit and the derived parameters.  

\begin{figure}[h!]
\includegraphics[width=1.0 \linewidth]{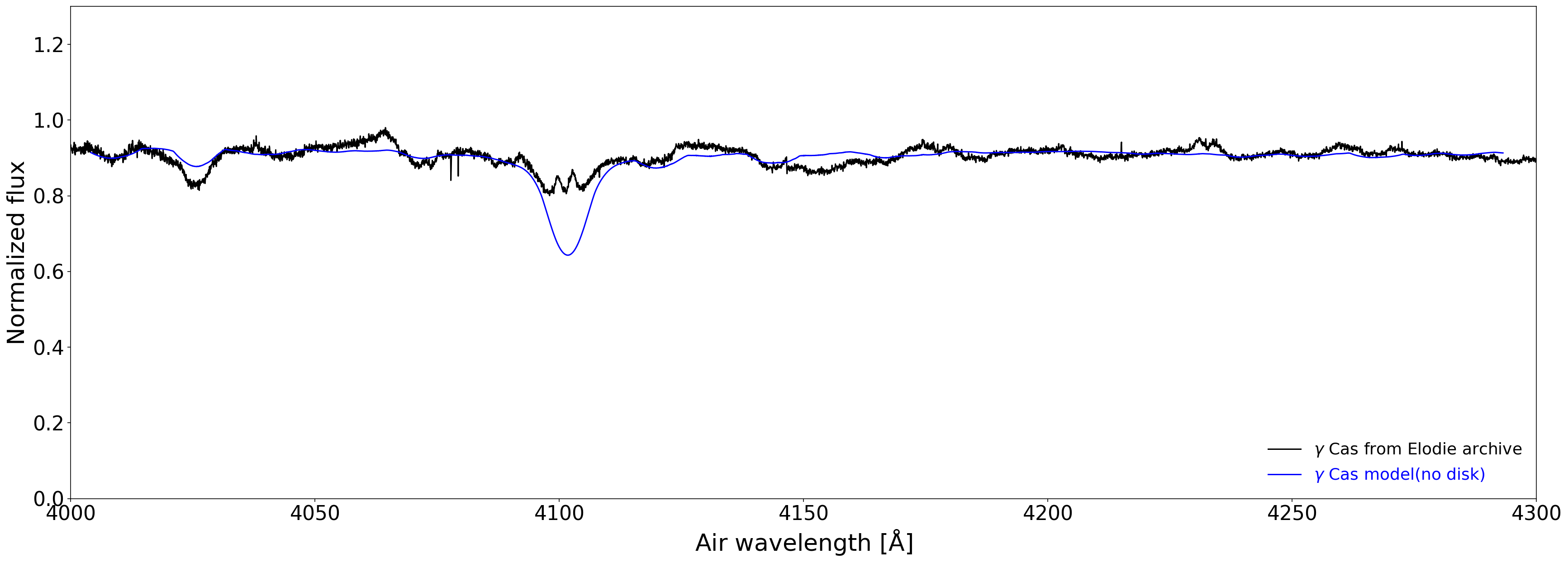}
    \caption{A normalized, high-resolution spectrum of $\gamma$ Cas in the VSII bandpass  from the ELODIE archive \citep[observation 20030816/0030, in black]{elodie_2004}
     and a normalized model spectrum (in blue) from the same model Roche-von Zeipel that best fits the VERITAS data in Table~\ref{tab:alldata}. The rotationally broadened H$\delta$ line
     at 4100~\AA\xspace 
     is filled in the emission from the disk surrounding $\gamma$ Cas. The disk is not included in our model.}
    \label{fig:high_res_spec}
\end{figure}

Our adopted polar gravity and inclination yield a projected rotation velocity
of $v\sin i = 389 \pm 20~{\rm km\, s^{-1}}$. \citet{vsini_is_good_match} 
found $v\sin i$ values 420-425 ${\rm km\,s^{-1}}$ by comparing models to the He I $\lambda$\,4471 line. 
Figure~\ref{fig:high_res_spec} shows a comparison between the best-fit model to
the squared visibility data and a high-resolution spectrum in the VSII band around 416 nm. Although the H$\delta$ line is filled with emission from the disk, the model roughly matches the wings of the line profile.  The other recognizable absorption line is He I $\lambda$\,4026, which is stronger than the model with roughly the same width.

\begin{deluxetable*}{lCl}[t]
\label{tab:stellar_parameters}
\tablecaption{Stellar Parameters for best fitting Roche-von Zeipel model for $\gamma$ Cas }
\tablecolumns{3}
\tablewidth{0pt}
\tablehead{
\colhead{Parameter} &
\colhead{Value} & 
\colhead{Reference}}
\startdata
\hline
\multicolumn{3}{c}{With a fixed inclination of the rotation axis, $i = 60^\circ$ and gravity darkening $\beta$ = 0.2}\\
\hline
Equatorial angular diameter, $\theta_{\rm eq}$ (mas)    &0.605$^{+0.041}_{-0.034}$  &This paper\\
Fraction of critical rotation rate, $\omega = \Omega/\Omega_{c}$ &0.990$^{+0.007}_{-0.013}$ &This paper\\
Parallax, $\varpi$ (mas)                          &5.94$\pm$0.12 &\citet{V07}\\
Equatorial radius, $R_{\rm eq}$  ($R_\sun$)  &10.9$^{+0.8}_{-0.6}$  & derived, $R\rm_{eq} = 107.48\frac{\theta_{eq}}{\varpi}$ \\
Polar Radius, $R_{\rm pole}$ ($R_\sun$)      &7.9$\pm0.4$ & derived, $R_{\rm pole} = \omega  R_{\rm equ}/\eta$ \\
\hline
\multicolumn{3}{c}{With a fixed $\log(g)_{\rm polar}$ = 3.82, $i = 60^\circ$, gravity darkening $\beta$ = 0.2}\\
\hline
Mass, $M$ ($M_\sun$)                             &15$\pm$2 &$ M = g_{\rm pole} R^2_{\rm pole}/{G}$\\ 
Equatorial velocity, $v_{\rm eq}$ (km\,s$^{-1}$),  &450$\pm$20 &$v_{\rm eq} = R_{\rm eq}\Omega$\\
Projected equatorial velocity, $v\sin\,i$ (${\rm km\,s^{-1}}$) &389$\pm$ 20 &derived from $v_{\rm eq}$ and $i = 60^\circ$, see  Figure~\ref{fig:high_res_spec}\\
Polar effective temperature, $T_{\rm pole}$ (K)  &$\simeq$ 26500 -28000 & spectrophotometry, see Figure~\ref{fig:SED_comparison} \\
Pole-equator difference, $\Delta T_{\rm eff}$ (K)  &$\simeq$ 9200 -12000   &derived, $\Delta T_{\rm eff} =  T_{\rm pole}\biggl[1-(\frac{\omega^2}{\eta^2}-\frac{8}{27}\eta\omega)^{\beta}\biggr]$
\enddata
\tablecomments{The parameters $R_{\rm pole}$ and $\Delta T_{\rm eff}$ depend on  $\eta = 3\cos\bigl[\frac{\pi + cos^{-1}(\omega)}{3}\bigr]$,
while the equatorial velocity, $v_{\rm eq}$, depends on  $\Omega = \omega\sqrt{\frac{8}{27}\frac{GM}{R_{\rm pole}^3}}$} 

\end{deluxetable*}

\section{Summary and Outlook}
\label{sec:Summary}

We have used the VERITAS telescope array to perform intensity interferometry measurements
  on the rapid rotator $\gamma$ Cassiopeiae, combining data from more than 160 pair-hours of observation.
While the size and elongation of the decretion disk of this star have been well measured at infrared
  wavelengths, measurements of the photosphere at visible wavelengths have previously only
  been sensitive to its size.
The six telescope pairs of the VSII observatory provide good coverage of the Fourier $u-v$ plane.
By varying the orientation as well as the magnitude of the baseline between telescopes,
we were able to measure the size, the equatorial bulge, and the orientation
  of the axis of rotation of the photosphere.
This is the first time that intensity interferometry has been used to measure the oblateness
  of a star.

A uniform ellipse fit of our data indicates that the angular size of the minor axis
  of the photosphere is $0.43\pm0.02\pm0.02$~mas, with a major-to-minor axis ratio of $1.28\pm0.04 \pm 0.02$.
The position angle of the minor axis, which corresponds to the 
rotator axis,
  is $116^\circ\pm5^\circ\pm7^\circ$, in agreement with earlier measurements of the accretion disk.

We have also compared our data to a Roche-von Zeipel model of a rapid rotator, which includes effects
  of limb and gravitational darkening as well as temperature gradients.
The temperature profile was adjusted to match published spectrophotometry measurements.
Our data is well fit within this model with an equatorial diameter of $0.604^{+0.041}_{-0.034}$~mas, position angle of $114.7^{+6.4}_{-5.7}$ degrees, and a lower limit on the angular velocity very near the critical value, $\Omega/\Omega_c=0.977$.
The equatorial radius is 10.9$^{+0.8}_{-0.6}\,\rm R_\odot$  and the mass of the star in the model is $15 \pm2\ M_\odot$.  

Previous stellar intensity interferometry measurements have been limited to extracting the overall size of the photosphere and resolving the extended decretion disk in its H$\alpha$ light \citep{Matthews2023}.
In order to take advantage of our coverage in the $u-v$ plane and gain sensitivity to the
  stellar shape, we found it important to remove artifacts from external radio interference
  and manual tracking corrections.
Our remaining systematic uncertainties are dominated by uncertainties in the clock start synchronization,
  the peak-finding algorithm, and by residual correlated noise in the background.
The overall systematic error is on the order of the statistical error, suggesting that only adding hours at previously unobserved hour angles would increase the precision of our measurement.

Having established the capability of stellar intensity interferometry to measure not only the size,
  but also the shape and orientation of a photosphere, we are continuing analysis of other rapid
  rotators, to expand the scope of data in the stellar catalog of bright stars.
This will provide valuable input and constraints to stellar models that must match both the
  spectral and the geometrical consequences of extreme rotation.

\section*{Acknowledgments}
This research is supported by grants from the U.S. Department of Energy Office of Science, the U.S. National Science Foundation and the Smithsonian Institution, by NSERC in Canada, and by the Helmholtz Association in Germany. We acknowledge the excellent work of the technical support staff at the Fred Lawrence Whipple Observatory and the collaborating institutions in the construction and operation of the VERITAS and VSII instruments. 
The authors gratefully acknowledge support under NSF grants \#AST 1806262, \#PHY 2117641, and \#PHY 2111531 for the construction and operation of the VSII instrumentation. 
We acknowledge the support of the Ohio Supercomputer Center and of an Ohio State University
College of Arts and Sciences Exploration Grant.
We acknowledge the financial support of the French National Research Agency (project I2C, ANR-20-CE31-0003).
This work made use of the High Performance Cluster ``Vega'' at Embry-Riddle Aeronautical University.  
This research has made use of the VizieR catalogue access tool, CDS,  Strasbourg, France (DOI : 10.26093/cds/vizier). 
The original description of the VizieR service was published in 2000, A\&AS 143, 23. 
Part of this research is based on INES data from the IUE satellite.

\software{PHOENIX (version 20.01.02B)~\citep{H99},
Astropy~\citep{Astropy},
specutils~\citep{specutils},
Minuit~\citep{James:1975dr},
root~\citep{Brun:1997pa},
Cobaya ~\citep{cobaya2021}, 
SciPy ~\citep{2020SciPy-NMeth}, Corner~\citep{corner}.}

\clearpage
\appendix
\restartappendixnumbering 

\section{Calculating uncertainties on peak areas with correlated background contribution}
\label{sec:appendix errorbars}

\begin{figure}[ht!]
    \centering
    \includegraphics[width=0.7\linewidth]{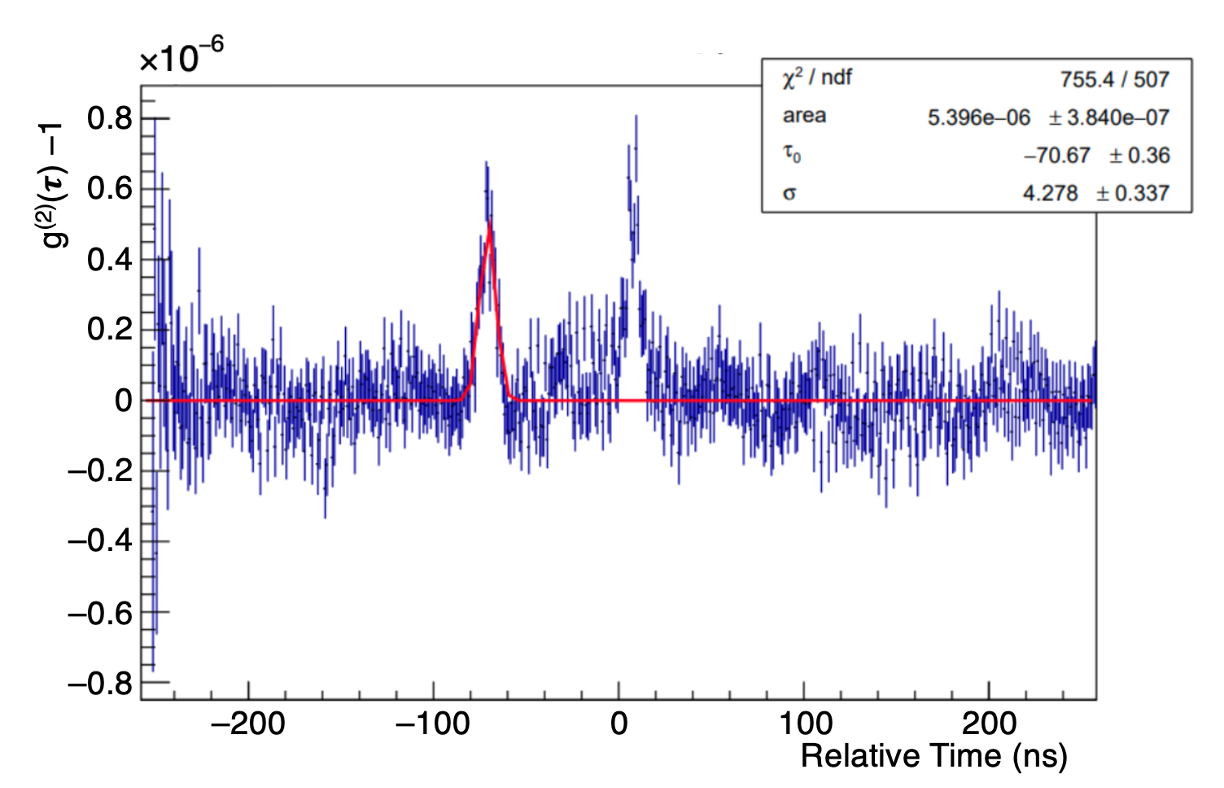}
    \caption{An example of one simulated peak inserted into the background noise and fitted with a Gaussian function to extract its area. }
    \label{fig:injextpeak}
\end{figure}

\begin{figure}[ht!]
    \centering
    \includegraphics[width=0.5\linewidth]{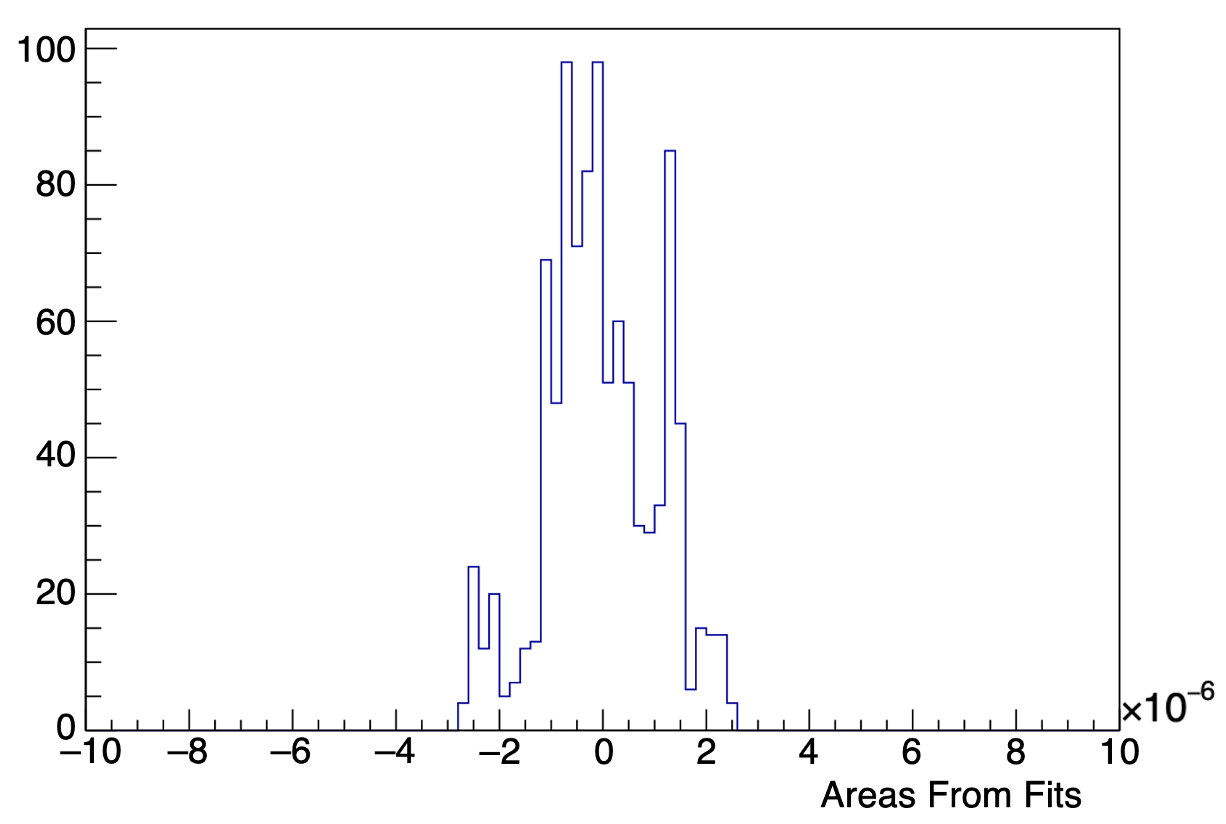}
    \caption{A distribution of the difference between the original known area of the simulated peaks and the areas found by the fitting algorithm once they are inserted into the background noise, for many fits. The 
    RMS of this distribution is the uncertainty on the peak.}
    \label{fig:newerrbardist}
\end{figure}

This appendix illustrates in more detail the calculation of an uncertainty on the area of the Gaussian peak, which we discuss in section \ref{sec:VisibilitiesAndUncertainties}. 
Figure \ref{fig:injextpeak} shows an example of a fit to a simulated peak located at about -70~ns in relative time, with the same area as the original correlation peak located at 0~ns in relative time. The simulated peak is inserted into the background noise away from the peak, fitted with the same procedure as the original peak, and subjected to the same quality cuts. This is repeated about 5000 times for each correlation function. The differences between the areas found from these fits and the true area of the simulated peak are added to a distribution like the one shown in figure \ref{fig:newerrbardist}. The uncertainty on the peak area is the RMS 
of this distribution, which describes how much the background noise can distort the area.

\section{Unnormalized visibilities above threshold}

Table~\ref{tab:alldata} lists all Gaussian integrals for visibilities
above thresholds discussed in section~\ref{sec:Analysis}.

\begin{deluxetable*}{ccccccc}[b]
\label{tab:alldata}
\tablecaption{$\gamma$ Cassiopeiae squared visibilities}
\tablecolumns{7}
\tablewidth{0pt}
\tablehead{
\colhead{UTC Date (yyyy-mm-dd hh:mm)} &
\colhead{Pair} &
\colhead{Area (fs)} &
\colhead{Baseline (m)} &
\colhead{u$\lambda$ (m)} & 
\colhead{v$\lambda$ (m)} &
\colhead{Hour angle (h)}}
\startdata
2023-12-25 01:57 & T2T4 & 7.2 $\pm$ 1.0 & 95.5 & -82.5 & 47.6 & 0.28 \\
2023-12-25 03:04 & T2T4 & 6.8 $\pm$ 1.3 & 90.0 & -86.2 & 25.0 & 1.44 \\
2023-12-26 01:34 & T1T2 & 7.4 $\pm$ 1.3 & 96.2 & -90.4 & -32.4 & 0.00 \\
2023-12-26 01:34 & T1T2 & 4.6 $\pm$ 0.9 & 98.5 & -83.5 & -51.9 & 0.98 \\
2023-12-26 01:34 & T1T3 & 5.1 $\pm$ 0.5 & 120.0 & -108.9 & 48.2 & 0.50 \\
2023-12-26 01:34 & T2T3 & 6.2 $\pm$ 0.8 & 94.3 & -15.5 & 92.9 & 0.00 \\
2023-12-26 01:34 & T2T3 & 7.0 $\pm$ 0.9 & 92.5 & -28.4 & 88.0 & 0.98 \\
2023-12-26 01:34 & T2T4 & 6.3 $\pm$ 1.0 & 96.5 & -80.5 & 52.8 & 0.00 \\
2023-12-26 01:34 & T2T4 & 5.7 $\pm$ 1.1 & 92.5 & -85.7 & 34.3 & 0.97 \\
2023-12-26 01:34 & T3T4 & 6.6 $\pm$ 1.1 & 76.5 & -65.0 & -40.1 & 0.01 \\
2023-12-26 01:34 & T3T4 & 5.6 $\pm$ 1.2 & 78.8 & -57.5 & -53.7 & 0.97 \\
2023-12-26 03:37 & T2T3 & 5.8 $\pm$ 1.5 & 89.2 & -40.1 & 79.6 & 2.04 \\
2023-12-26 03:37 & T2T3 & 6.5 $\pm$ 1.6 & 84.7 & -48.5 & 69.3 & 3.05 \\
2023-12-26 03:37 & T2T4 & 7.5 $\pm$ 1.1 & 86.6 & -85.3 & 14.3 & 1.99 \\
2023-12-26 03:37 & T2T4 & 8.5 $\pm$ 1.5 & 78.4 & -78.0 & -6.6 & 3.10 \\
2023-12-26 03:37 & T3T4 & 7.0 $\pm$ 1.3 & 80.2 & -45.7 & -65.8 & 1.99 \\
2023-12-26 03:37 & T3T4 & 3.8 $\pm$ 0.9 & 80.9 & -29.7 & -75.1 & 3.06 \\
2023-12-26 05:38 & T1T2 & 4.0 $\pm$ 1.4 & 99.4 & -30.3 & -94.4 & 4.07 \\
2023-12-26 05:38 & T1T2 & 4.4 $\pm$ 1.7 & 99.2 & -7.5 & -98.6 & 5.04 \\
\enddata
\end{deluxetable*}
\begin{deluxetable*}{ccccccc}[b]
\startdata
2023-12-26 05:38 & T1T3 & 5.7 $\pm$ 0.8 & 92.0 & -83.9 & -37.1 & 4.07 \\
2023-12-26 05:38 & T2T3 & 5.4 $\pm$ 1.1 & 78.6 & -53.6 & 57.3 & 4.07 \\
2023-12-26 05:38 & T2T3 & 7.9 $\pm$ 1.4 & 71.2 & -54.9 & 45.2 & 5.04 \\
2023-12-26 05:38 & T2T4 & 9.1 $\pm$ 1.0 & 70.1 & -66.2 & -22.6 & 4.07 \\
2023-12-26 05:38 & T2T4 & 8.7 $\pm$ 1.1 & 61.7 & -50.2 & -35.5 & 5.03 \\
2023-12-26 05:38 & T3T4 & 5.4 $\pm$ 0.9 & 81.1 & -12.6 & -80.0 & 4.07 \\
2023-12-26 05:38 & T3T4 & 8.6 $\pm$ 1.0 & 81.1 & 4.9 & -80.8 & 5.04 \\
2023-12-26 07:39 & T1T2 & 4.5 $\pm$ 1.9 & 99.3 & 20.3 & -96.7 & 6.21 \\
2023-12-26 07:39 & T1T3 & 4.4 $\pm$ 1.1 & 73.8 & -31.3 & -65.8 & 6.21 \\
2023-12-26 07:39 & T2T3 & 12.2 $\pm$ 2.2 & 60.3 & -51.7 & 31.0 & 6.20 \\
2023-12-26 07:39 & T2T4 & 8.6 $\pm$ 1.5 & 53.6 & -26.6 & -45.7 & 6.21 \\
2023-12-26 07:39 & T3T4 & 6.3 $\pm$ 0.9 & 81.0 & 25.1 & -76.6 & 6.21 \\
2023-12-27 01:43 & T1T2 & 5.0 $\pm$ 1.2 & 96.8 & -89.4 & -36.5 & 0.20 \\
2023-12-27 01:43 & T1T3 & 3.6 $\pm$ 0.8 & 119.2 & -109.8 & 44.2 & 0.66 \\
2023-12-27 01:43 & T2T3 & 6.4 $\pm$ 0.9 & 91.9 & -31.0 & 86.5 & 1.20 \\
2023-12-27 01:43 & T2T4 & 7.5 $\pm$ 1.5 & 95.8 & -82.0 & 49.1 & 0.20 \\
2023-12-27 01:43 & T2T4 & 10.8 $\pm$ 1.4 & 91.4 & -86.1 & 30.1 & 1.18 \\
2023-12-27 01:43 & T3T4 & 6.0 $\pm$ 1.2 & 77.1 & -63.8 & -43.1 & 0.21 \\
2023-12-27 03:44 & T1T2 & 4.6 $\pm$ 0.9 & 99.7 & -66.7 & -73.7 & 2.24 \\
2023-12-27 03:44 & T1T3 & 6.2 $\pm$ 1.1 & 105.4 & -104.3 & -5.8 & 2.65 \\
2023-12-27 03:44 & T2T3 & 8.0 $\pm$ 1.2 & 88.5 & -42.0 & 77.7 & 2.24 \\
2023-12-27 03:44 & T2T3 & 7.9 $\pm$ 1.7 & 83.7 & -49.7 & 67.2 & 3.24 \\
2023-12-27 03:44 & T2T4 & 11.0 $\pm$ 1.5 & 84.9 & -84.2 & 9.4 & 2.24 \\
2023-12-27 03:44 & T2T4 & 8.4 $\pm$ 2.1 & 77.3 & -76.6 & -8.8 & 3.22 \\
2023-12-27 03:44 & T3T4 & 7.0 $\pm$ 1.4 & 80.4 & -42.2 & -68.3 & 2.24 \\
2023-12-27 05:47 & T1T2 & 3.7 $\pm$ 1.6 & 99.3 & -25.8 & -95.6 & 4.26 \\
2023-12-27 05:47 & T1T3 & 6.9 $\pm$ 1.0 & 90.0 & -79.9 & -40.7 & 4.27 \\
2023-12-27 05:47 & T1T3 & 9.6 $\pm$ 1.7 & 81.6 & -60.2 & -54.8 & 5.14 \\
\enddata
\end{deluxetable*}
\begin{deluxetable*}{ccccccc}[b]
\startdata
2023-12-27 05:47 & T2T3 & 11.1 $\pm$ 1.5 & 70.5 & -54.9 & 44.1 & 5.13 \\
2023-12-27 05:47 & T2T4 & 7.4 $\pm$ 1.3 & 68.4 & -63.2 & -25.5 & 4.27 \\
2023-12-27 05:47 & T2T4 & 8.5 $\pm$ 1.8 & 60.7 & -47.9 & -37.0 & 5.16 \\
2023-12-27 05:47 & T3T4 & 7.1 $\pm$ 1.2 & 81.1 & 6.3 & -80.8 & 5.12 \\
2023-01-08 05:23 & T1T3 & 7.8 $\pm$ 1.3 & 86.1 & -71.5 & -47.5 & 4.66 \\
2023-01-08 05:23 & T1T3 & 7.7 $\pm$ 1.1 & 76.9 & -45.4 & -61.6 & 5.70 \\
2023-01-08 05:23 & T2T3 & 8.8 $\pm$ 1.3 & 74.4 & -54.8 & 50.3 & 4.64 \\
2023-01-08 05:23 & T2T3 & 9.3 $\pm$ 1.3 & 65.7 & -53.9 & 37.5 & 5.66 \\
2023-01-08 05:23 & T2T4 & 7.5 $\pm$ 1.2 & 64.9 & -56.7 & -31.0 & 4.67 \\
2023-01-08 05:23 & T2T4 & 9.3 $\pm$ 1.8 & 56.7 & -37.7 & -42.0 & 5.68 \\
2023-01-08 05:23 & T3T4 & 6.4 $\pm$ 1.2 & 81.1 & -1.9 & -81.0 & 4.66 \\
2023-02-02 03:10 & T1T2 & 5.1 $\pm$ 1.3 & 99.4 & -29.8 & -94.6 & 4.09 \\
2023-02-02 03:10 & T1T3 & 8.8 $\pm$ 1.5 & 91.7 & -83.4 & -37.6 & 4.10 \\
2023-02-02 03:10 & T1T3 & 7.5 $\pm$ 1.1 & 82.2 & -61.7 & -53.8 & 5.07 \\
2023-02-02 03:10 & T2T3 & 9.1 $\pm$ 0.9 & 78.4 & -53.7 & 57.1 & 4.09 \\
2023-02-02 03:10 & T2T3 & 7.6 $\pm$ 1.7 & 71.0 & -54.9 & 44.9 & 5.07 \\
2023-02-02 03:10 & T2T4 & 6.0 $\pm$ 1.6 & 70.0 & -65.9 & -22.9 & 4.08 \\
2023-02-02 03:10 & T2T4 & 11.7 $\pm$ 1.7 & 61.4 & -49.4 & -36.0 & 5.07 \\
2023-02-02 03:10 & T3T4 & 7.5 $\pm$ 1.7 & 81.1 & -12.3 & -80.0 & 4.08 \\
2023-02-02 03:10 & T3T4 & 7.3 $\pm$ 1.1 & 81.1 & 5.5 & -80.8 & 5.08 \\
2023-02-02 05:12 & T1T3 & 4.0 $\pm$ 1.2 & 73.6 & -31.0 & -66.0 & 6.22 \\
2023-02-02 05:12 & T2T3 & 11.2 $\pm$ 1.6 & 60.2 & -51.7 & 30.8 & 6.22 \\
2023-02-02 05:12 & T2T4 & 7.9 $\pm$ 1.4 & 53.5 & -26.5 & -45.8 & 6.21 \\
2024-02-19 02:31 & T1T3 & 6.1 $\pm$ 1.0 & 87.2 & -73.9 & -45.7 & 4.55 \\
2024-02-19 02:31 & T2T3 & 9.9 $\pm$ 1.1 & 75.1 & -54.7 & 51.3 & 4.55 \\
2024-02-19 02:31 & T3T4 & 4.7 $\pm$ 1.3 & 81.1 & -4.1 & -80.9 & 4.54 \\
2024-02-19 03:38 & T1T2 & 4.4 $\pm$ 1.7 & 99.2 & 7.9 & -98.6 & 5.68 \\
2024-02-19 03:38 & T1T3 & 5.7 $\pm$ 1.6 & 77.4 & -46.9 & -61.0 & 5.65 \\
\enddata
\end{deluxetable*}
\begin{deluxetable*}{ccccccc}
\startdata
2024-02-19 03:38 & T1T4 & 4.1 $\pm$ 1.1 & 144.4 & -31.1 & -140.4 & 5.65 \\
2024-02-19 03:38 & T2T3 & 9.9 $\pm$ 1.5 & 65.5 & -53.8 & 37.3 & 5.68 \\
2024-02-19 03:38 & T2T4 & 10.6 $\pm$ 2.1 & 56.7 & -37.5 & -42.1 & 5.68 \\
2024-02-20 02:07 & T1T2 & 3.7 $\pm$ 1.9 & 99.4 & -29.9 & -94.6 & 4.09 \\
2024-02-20 02:07 & T1T3 & 9.0 $\pm$ 2.3 & 91.8 & -83.6 & -37.5 & 4.09 \\
2024-02-20 02:07 & T2T3 & 9.4 $\pm$ 1.9 & 78.6 & -53.7 & 57.4 & 4.07 \\
2024-02-20 02:07 & T2T4 & 4.4 $\pm$ 1.6 & 70.1 & -66.1 & -22.7 & 4.07 \\
2024-02-20 03:14 & T1T3 & 6.1 $\pm$ 1.7 & 81.3 & -59.4 & -55.3 & 5.17 \\
2024-02-20 03:14 & T2T3 & 7.4 $\pm$ 2.8 & 70.2 & -54.9 & 43.7 & 5.16 \\
2024-02-21 02:22 & T1T3 & 10.3 $\pm$ 1.1 & 88.1 & -75.8 & -44.2 & 4.46 \\
2024-02-21 02:22 & T2T3 & 9.6 $\pm$ 2.9 & 75.8 & -54.6 & 52.5 & 4.46 \\
2024-02-22 03:26 & T1T3 & 5.8 $\pm$ 2.0 & 78.9 & -52.4 & -58.9 & 5.44 \\
2024-02-22 03:26 & T2T4 & 7.4 $\pm$ 1.6 & 58.5 & -42.7 & -39.9 & 5.43 \\
2024-02-22 03:26 & T3T4 & 4.5 $\pm$ 1.1 & 81.1 & 11.9 & -80.2 & 5.44 \\
2024-02-22 03:59 & T2T3 & 10.2 $\pm$ 2.3 & 62.7 & -52.9 & 33.5 & 5.98 \\
2024-02-22 03:59 & T3T4 & 7.9 $\pm$ 2.3 & 81.0 & 21.4 & -78.1 & 5.98 \\
2024-02-22 04:32 & T1T3 & 5.6 $\pm$ 2.1 & 71.5 & -18.8 & -68.7 & 6.65 \\
2024-02-22 04:32 & T2T3 & 9.0 $\pm$ 1.8 & 55.7 & -49.4 & 25.7 & 6.65 \\
2024-02-22 04:32 & T2T4 & 8.8 $\pm$ 1.4 & 51.4 & -17.2 & -48.2 & 6.64 \\
2024-05-18 10:00 & T1T2 & 6.7 $\pm$ 1.7 & 62.4 & -14.9 & 60.2 & -6.02 \\
2024-05-18 10:00 & T2T3 & 8.2 $\pm$ 1.3 & 80.1 & 52.8 & 60.1 & -6.02 \\
2024-05-18 10:00 & T3T4 & 16.8 $\pm$ 1.3 & 42.0 & -21.4 & 35.8 & -6.02 \\
2024-05-19 10:07 & T1T2 & 7.3 $\pm$ 1.4 & 62.8 & -18.4 & 59.6 & -5.87 \\
2024-05-19 10:07 & T2T3 & 4.7 $\pm$ 1.4 & 81.0 & 52.2 & 61.9 & -5.87 \\
2024-05-19 10:07 & T3T4 & 11.9 $\pm$ 1.1 & 42.6 & -23.7 & 35.1 & -5.88 \\
2024-12-13 01:34 & T1T2 & 4.8 $\pm$ 1.2 & 93.3 & -91.6 & -16.6 & -0.75 \\
2024-12-13 01:34 & T2T3 & 6.8 $\pm$ 0.9 & 94.9 & -4.1 & 94.7 & -0.80 \\
2024-12-13 01:34 & T2T4 & 6.5 $\pm$ 1.3 & 98.6 & -72.3 & 66.8 & -0.80 \\
2024-12-13 01:34 & T3T4 & 8.5 $\pm$ 1.0 & 73.7 & -68.1 & -27.8 & -0.80 \\
\enddata
\end{deluxetable*}
\begin{deluxetable*}{ccccccc}
\startdata
2024-12-13 02:40 & T1T2 & 4.1 $\pm$ 1.0 & 97.0 & -88.9 & -38.3 & 0.29 \\
2024-12-13 02:40 & T1T3 & 4.8 $\pm$ 1.1 & 121.1 & -108.4 & 53.4 & 0.29 \\
2024-12-13 02:40 & T2T3 & 6.7 $\pm$ 0.9 & 93.8 & -19.5 & 91.7 & 0.30 \\
2024-12-13 02:40 & T2T4 & 6.3 $\pm$ 1.3 & 95.4 & -82.6 & 47.4 & 0.29 \\
2024-12-13 02:40 & T3T4 & 5.8 $\pm$ 1.3 & 77.3 & -63.2 & -44.4 & 0.29 \\
2024-12-15 05:43 & T1T2 & 4.0 $\pm$ 1.0 & 99.6 & -44.7 & -88.7 & 3.41 \\
2024-12-15 05:43 & T1T3 & 6.8 $\pm$ 1.3 & 98.6 & -95.6 & -23.4 & 3.40 \\
2024-12-15 05:43 & T2T3 & 5.4 $\pm$ 1.1 & 82.7 & -50.8 & 65.2 & 3.41 \\
2024-12-15 05:43 & T3T4 & 7.3 $\pm$ 1.0 & 81.0 & -23.8 & -77.2 & 3.41 \\
2024-12-17 02:28 & T1T3 & 3.8 $\pm$ 0.9 & 120.8 & -108.8 & 51.9 & 0.35 \\
2024-12-17 02:28 & T2T3 & 7.1 $\pm$ 0.6 & 93.7 & -20.3 & 91.4 & 0.36 \\
2024-12-17 02:28 & T2T4 & 6.9 $\pm$ 0.7 & 95.2 & -83.0 & 46.2 & 0.36 \\
2024-12-17 02:28 & T3T4 & 7.4 $\pm$ 1.0 & 77.5 & -62.7 & -45.2 & 0.36 \\
\enddata
\end{deluxetable*}
\clearpage

\bibliography{references}{}

\begin{thebibliography}{}
\expandafter\ifx\csname natexlab\endcsname\relax\def\natexlab#1{#1}\fi
\providecommand{\url}[1]{\href{#1}{#1}}
\providecommand{\dodoi}[1]{doi:~\href{http://doi.org/#1}{\nolinkurl{#1}}}
\providecommand{\doeprint}[1]{\href{http://ascl.net/#1}{\nolinkurl{http://ascl.net/#1}}}
\providecommand{\doarXiv}[1]{\href{https://arxiv.org/abs/#1}{\nolinkurl{https://arxiv.org/abs/#1}}}

\bibitem[{Abe {et~al.}(2024)Abe, Abhir, Acciari, Aguasca-Cabot, Agudo, Aniello, Ansoldi, Antonelli, Arbet Engels, Arcaro, Artero, Asano, Babić, Baquero, de Almeida, Barrio, Batković, Bautista, Baxter, González, Bernardini, Bernardos, Bernete, Berti, Besenrieder, Bigongiari, Biland, Blanch, Bonnoli, Bošnjak, Burelli, Busetto, Campoy-Ordaz, Carosi, Carosi, Carretero-Castrillo, Ceribella, Chai, Cifuentes, Colombo, Contreras, Cortina, Covino, D’Amico, D’Elia, Da Vela, Dazzi, De Angelis, De Lotto, de Menezes, Del~Popolo, Delfino, Delgado, Mendez, Pierro, Venere, Prester, Donini, Dorner, Doro, Elsaesser, Emery, Escudero, Fariña, Fattorini, Foffano, Font, Fröse, Fukami, Fukazawa, López, Garczarczyk, Gasparyan, Gaug, Paiva, Giglietto, Giordano, Gliwny, Gradetzke, Grau, Green, Green, Günther, Hadasch, Hahn, Hassan, Heckmann, Herrera, Hrupec, Hütten, Imazawa, Ishio, Jiménez Martínez, Jormanainen, Kayanoki, Kerszberg, Kluge, Kobayashi, Kouch, Kubo, Kushida, Láinez, Lamastra, Leone, Lindfors,
  Linhoff, Lombardi, Longo, López-Coto, López-Moya, López-Oramas, Loporchio, Lorini, Lyard, Fraga, Majumdar, Makariev, Maneva, Mang, Manganaro, Mangano, Mannheim, Mariotti, Martínez, Martínez-Chicharro, Mas-Aguilar, Mazin, Menchiari, Mender, Miceli, Miener, Miranda, Mirzoyan, González, Molina, Mondal, Moralejo, Morcuende, Nakamori, Nanci, Neustroev, Nickel, Rosillo, Nigro, Nikolić, Nilsson, Nishijima, Ekoume, Noda, Nozaki, Ohtani, Okumura, Otero-Santos, Paiano, Palatiello, Paneque, Paoletti, Paredes, Peresano, Persic, Pihet, Pirola, Podobnik, Moroni, Prandini, Principe, Priyadarshi, Rhode, Ribó, Rico, Righi, Sahakyan, Saito, Satalecka, Saturni, Schleicher, Schmidt, Schmuckermaier, Schubert, Schweizer, Sciaccaluga, Silvestri, Sitarek, Sliusar, Sobczynska, Spolon, Stamerra, Strišković, Strom, Strzys, Suda, Surić, Suutarinen, Tajima, Takahashi, Takeishi, Temnikov, Terauchi, Terzić, Teshima, Truzzi, Tutone, Ubach, van Scherpenberg, Acosta, Ventura, Viale, Vigorito, Vitale, Walter, Will, Wunderlich,
  Yamamoto, Chon, Díaz, Fiori, Lobo, Naletto, Polo, Rodríguez-Vázquez, Saha, \& Zampieri}]{10.1093/mnras/stae697}
Abe, S., Abhir, J., Acciari, V.~A., {et~al.} 2024, Monthly Notices of the Royal Astronomical Society, 529, 4387, \dodoi{10.1093/mnras/stae697}

\bibitem[{Abeysekara {et~al.}(2020)Abeysekara, Benbow, Brill, Buckley, Christiansen, Chromey, Daniel, Davis, Falcone, Feng, Finley, Fortson, Furniss, Gent, Giuri, Gueta, Hanna, Hassan, Hervet, Holder, Hughes, Humensky, Kaaret, Kertzman, Kieda, Krennrich, Kumar, LeBohec, Lin, Lundy, Maier, Matthews, Moriarty, Mukherjee, Nievas-Rosillo, O’Brien, Ong, Otte, Pfrang, Pohl, Prado, Pueschel, Quinn, Ragan, Reynolds, Ribeiro, Richards, Roache, Ryan, Santander, Sembroski, Wakely, Weinstein, Wilcox, Williams, \& Williamson}]{natureSIIDemo}
Abeysekara, A.~U., Benbow, W., Brill, A., {et~al.} 2020, Nature Astronomy, 4, 1164–1169, \dodoi{10.1038/s41550-020-1143-y}

\bibitem[{Acharyya {et~al.}(2024)}]{VERITAS:2024quv}
Acharyya, A., {et~al.} 2024, Astrophys. J., 966, 28, \dodoi{10.3847/1538-4357/ad2b68}

\bibitem[{{Astropy Collaboration} {et~al.}(2018){Astropy Collaboration}, {Price-Whelan}, {Sip{\H{o}}cz}, {G{\"u}nther}, {Lim}, {Crawford}, {Conseil}, {Shupe}, {Craig}, {Dencheva}, {Ginsburg}, {VanderPlas}, {Bradley}, {P{\'e}rez-Su{\'a}rez}, {de Val-Borro}, {Aldcroft}, {Cruz}, {Robitaille}, {Tollerud}, {Ardelean}, {Babej}, {Bach}, {Bachetti}, {Bakanov}, {Bamford}, {Barentsen}, {Barmby}, {Baumbach}, {Berry}, {Biscani}, {Boquien}, {Bostroem}, {Bouma}, {Brammer}, {Bray}, {Breytenbach}, {Buddelmeijer}, {Burke}, {Calderone}, {Cano Rodr{\'\i}guez}, {Cara}, {Cardoso}, {Cheedella}, {Copin}, {Corrales}, {Crichton}, {D'Avella}, {Deil}, {Depagne}, {Dietrich}, {Donath}, {Droettboom}, {Earl}, {Erben}, {Fabbro}, {Ferreira}, {Finethy}, {Fox}, {Garrison}, {Gibbons}, {Goldstein}, {Gommers}, {Greco}, {Greenfield}, {Groener}, {Grollier}, {Hagen}, {Hirst}, {Homeier}, {Horton}, {Hosseinzadeh}, {Hu}, {Hunkeler}, {Ivezi{\'c}}, {Jain}, {Jenness}, {Kanarek}, {Kendrew}, {Kern}, {Kerzendorf}, {Khvalko}, {King}, {Kirkby}, {Kulkarni},
  {Kumar}, {Lee}, {Lenz}, {Littlefair}, {Ma}, {Macleod}, {Mastropietro}, {McCully}, {Montagnac}, {Morris}, {Mueller}, {Mumford}, {Muna}, {Murphy}, {Nelson}, {Nguyen}, {Ninan}, {N{\"o}the}, {Ogaz}, {Oh}, {Parejko}, {Parley}, {Pascual}, {Patil}, {Patil}, {Plunkett}, {Prochaska}, {Rastogi}, {Reddy Janga}, {Sabater}, {Sakurikar}, {Seifert}, {Sherbert}, {Sherwood-Taylor}, {Shih}, {Sick}, {Silbiger}, {Singanamalla}, {Singer}, {Sladen}, {Sooley}, {Sornarajah}, {Streicher}, {Teuben}, {Thomas}, {Tremblay}, {Turner}, {Terr{\'o}n}, {van Kerkwijk}, {de la Vega}, {Watkins}, {Weaver}, {Whitmore}, {Woillez}, {Zabalza}, \& {Astropy Contributors}}]{Astropy}
{Astropy Collaboration}, {Price-Whelan}, A.~M., {Sip{\H{o}}cz}, B.~M., {et~al.} 2018, \aj, 156, 123, \dodoi{10.3847/1538-3881/aabc4f}

\bibitem[{{Aufdenberg} {et~al.}(2006){Aufdenberg}, {M{\'e}rand}, {Coud{\'e} du Foresto}, {Absil}, {Di Folco}, {Kervella}, {Ridgway}, {Berger}, {ten Brummelaar}, {McAlister}, {Sturmann}, {Sturmann}, \& {Turner}}]{vega06}
{Aufdenberg}, J.~P., {M{\'e}rand}, A., {Coud{\'e} du Foresto}, V., {et~al.} 2006, \apj, 645, 664, \dodoi{10.1086/504149}

\bibitem[{Brun \& Rademakers(1997)}]{Brun:1997pa}
Brun, R., \& Rademakers, F. 1997, Nucl. Instrum. Meth. A, 389, 81, \dodoi{10.1016/S0168-9002(97)00048-X}

\bibitem[{{Chauville} {et~al.}(2001){Chauville}, {Zorec}, {Ballereau}, {Morrell}, {Cidale}, \& {Garcia}}]{vsini_is_good_match}
{Chauville}, J., {Zorec}, J., {Ballereau}, D., {et~al.} 2001, \aap, 378, 861, \dodoi{10.1051/0004-6361:20011202}

\bibitem[{{Che} {et~al.}(2011){Che}, {Monnier}, {Zhao}, {Pedretti}, {Thureau}, {M{\'e}rand}, {ten Brummelaar}, {McAlister}, {Ridgway}, {Turner}, {Sturmann}, \& {Sturmann}}]{che2011}
{Che}, X., {Monnier}, J.~D., {Zhao}, M., {et~al.} 2011, \apj, 732, 68, \dodoi{10.1088/0004-637X/732/2/68}

\bibitem[{Dravins {et~al.}(2013)Dravins, LeBohec, Jensen, \& Nuñez}]{DRAVINS2013331}
Dravins, D., LeBohec, S., Jensen, H., \& Nuñez, P.~D. 2013, Astroparticle Physics, 43, 331, \dodoi{https://doi.org/10.1016/j.astropartphys.2012.04.017}

\bibitem[{Earl {et~al.}(2022)Earl, Tollerud, Jones, O'Steen, Kerzendorf, Busko, shaileshahuja, D'Avella, Robitaille, Ginsburg, Homeier, Sipőcz, Averbukh, Tocknell, Cherinka, Ogaz, Geda, Lim, Davies, Günther, Barbary, Foster, Conroy, Droettboom, Torres, Bray, Casey, Teuben, Crawford, \& Ferguson}]{specutils}
Earl, N., Tollerud, E., Jones, C., {et~al.} 2022, astropy/specutils: V1.7.0, v1.7.0,  Zenodo, \dodoi{10.5281/zenodo.6207491}

\bibitem[{Foreman-Mackey(2016)}]{corner}
Foreman-Mackey, D. 2016, The Journal of Open Source Software, 1, 24, \dodoi{10.21105/joss.00024}

\bibitem[{{Gonz{\'a}lez-Riestra} {et~al.}(2000){Gonz{\'a}lez-Riestra}, {Cassatella}, {Solano}, {Altamore}, \& {Wamsteker}}]{INES_IUE_data}
{Gonz{\'a}lez-Riestra}, R., {Cassatella}, A., {Solano}, E., {Altamore}, A., \& {Wamsteker}, W. 2000, \aaps, 141, 343, \dodoi{10.1051/aas:2000317}

\bibitem[{{Hanbury Brown}(1974)}]{1974RHB_book}
{Hanbury Brown}, R. 1974, {The intensity interferometer. Its applications to astronomy} (London: Taylor \& Francis Ltd.)

\bibitem[{{Hanbury Brown} \& Twiss(1954)}]{Brown01071954}
{Hanbury Brown}, R., \& Twiss, R. 1954, The London, Edinburgh, and Dublin Philosophical Magazine and Journal of Science, 45, 663, \dodoi{10.1080/14786440708520475}

\bibitem[{{Hauschildt} \& {Baron}(1999)}]{H99}
{Hauschildt}, P.~H., \& {Baron}, E. 1999, Journal of Computational and Applied Mathematics, 109, 41, \dodoi{10.48550/arXiv.astro-ph/9808182}

\bibitem[{Holder {et~al.}(2006)Holder, Atkins, Badran, Blaylock, Bradbury, Buckley, Byrum, Carter-Lewis, Celik, Chow, Cogan, Cui, Daniel, {de la Calle Perez}, Dowdall, Dowkontt, Duke, Falcone, Fegan, Finley, Fortin, Fortson, Gibbs, Gillanders, Glidewell, Grube, Gutierrez, Gyuk, Hall, Hanna, Hays, Horan, Hughes, Humensky, Imran, Jung, Kaaret, Kenny, Kieda, Kildea, Knapp, Krawczynski, Krennrich, Lang, LeBohec, Linton, Little, Maier, Manseri, Milovanovic, Moriarty, Mukherjee, Ogden, Ong, Petry, Perkins, Pizlo, Pohl, Quinn, Ragan, Reynolds, Roache, Rose, Schroedter, Sembroski, Sleege, Steele, Swordy, Syson, Toner, Valcarcel, Vassiliev, Wakely, Weekes, White, Williams, \& Wagner}]{HOLDER2006391}
Holder, J., Atkins, R., Badran, H., {et~al.} 2006, Astroparticle Physics, 25, 391, \dodoi{https://doi.org/10.1016/j.astropartphys.2006.04.002}

\bibitem[{James \& Roos(1975)}]{James:1975dr}
James, F., \& Roos, M. 1975, Comput. Phys. Commun., 10, 343, \dodoi{10.1016/0010-4655(75)90039-9}

\bibitem[{Kieda {et~al.}(2021)}]{VERITAS:2021xyw}
Kieda, D., {et~al.} 2021, PoS, ICRC2021, 803, \dodoi{10.22323/1.395.0803}

\bibitem[{{Lailey} \& {Sigut}(2024)}]{laily_sigut_2024}
{Lailey}, B.~D., \& {Sigut}, T.~A.~A. 2024, \mnras, 527, 2585, \dodoi{10.1093/mnras/stad3321}

\bibitem[{Matthews(2020)}]{NolanMatthewsThesis}
Matthews, N. 2020, PhD thesis, University of Utah

\bibitem[{{Matthews} {et~al.}(2023){Matthews}, {Rivet}, {Vernet}, {Hugbart}, {Labeyrie}, {Kaiser}, {Chab{\'e}}, {Courde}, {Lai}, {Vakili}, {Garde}, \& {Guerin}}]{Matthews2023}
{Matthews}, N., {Rivet}, J.-P., {Vernet}, D., {et~al.} 2023, \aj, 165, 117, \dodoi{10.3847/1538-3881/acb142}

\bibitem[{{Moultaka} {et~al.}(2004){Moultaka}, {Ilovaisky}, {Prugniel}, \& {Soubiran}}]{elodie_2004}
{Moultaka}, J., {Ilovaisky}, S.~A., {Prugniel}, P., \& {Soubiran}, C. 2004, \pasp, 116, 693, \dodoi{10.1086/422177}

\bibitem[{Rose(2025)}]{RoseThesis2025}
Rose, J. 2025, Bachelor's honors thesis, The Ohio State University.
\newblock \url{https://kb.osu.edu/handle/1811/105962}

\bibitem[{Rou {et~al.}(2013)Rou, Nuñez, Kieda, \& LeBohec}]{10.1093/mnras/stt123}
Rou, J., Nuñez, P.~D., Kieda, D., \& LeBohec, S. 2013, Monthly Notices of the Royal Astronomical Society, 430, 3187, \dodoi{10.1093/mnras/stt123}

\bibitem[{{Sackrider} \& {Aufdenberg}(2023)}]{sack23}
{Sackrider}, J.~L., \& {Aufdenberg}, J.~P. 2023, Research Notes of the American Astronomical Society, 7, 216, \dodoi{10.3847/2515-5172/ad023b}

\bibitem[{Scott(2023)}]{ScottThesis:2023}
Scott, M. 2023, Bachelor's honors thesis, The Ohio State University.
\newblock \url{http://hdl.handle.net/1811/102963}

\bibitem[{{Sigut} {et~al.}(2020){Sigut}, {Mahjour}, \& {Tycner}}]{sigut_2020}
{Sigut}, T.~A.~A., {Mahjour}, A.~K., \& {Tycner}, C. 2020, \apj, 894, 18, \dodoi{10.3847/1538-4357/ab8386}

\bibitem[{{Smith}(2019)}]{myron2019}
{Smith}, M.~A. 2019, \pasp, 131, 044201, \dodoi{10.1088/1538-3873/aaf70b}

\bibitem[{{Stee, Ph.} {et~al.}(2012){Stee, Ph.}, {Delaa, O.}, {Monnier, J. D.}, {Meilland, A.}, {Perraut, K.}, {Mourard, D.}, {Che, X.}, {Schaefer, G. H.}, {Pedretti, E.}, {Smith, M. A.}, {Lopes de Oliveira, R.}, {Motch, C.}, {Henry, G. W.}, {Richardson, N. D.}, {Bjorkman, K. S.}, {Bücke, R.}, {Pollmann, E.}, {Zorec, J.}, {Gies, D. R.}, {ten Brummelaar, T.}, {McAlister, H. A.}, {Turner, N. H.}, {Sturmann, J.}, {Sturmann, L.}, \& {Ridgway, S. T.}}]{refId0}
{Stee, Ph.}, {Delaa, O.}, {Monnier, J. D.}, {et~al.} 2012, \aap, 545, A59, \dodoi{10.1051/0004-6361/201219234}

\bibitem[{Thompson {et~al.}(2001)Thompson, Moran, \& Swenson~Jr.}]{thompson2001}
Thompson, A.~R., Moran, J.~M., \& Swenson~Jr., G.~W. 2001, Interferometry and Synthesis in Radio Astronomy (WILEY‐VCH Verlag GmbH \& Co. KGaA)

\bibitem[{{Torrado} \& {Lewis}(2021)}]{cobaya2021}
{Torrado}, J., \& {Lewis}, A. 2021, \jcap, 2021, 057, \dodoi{10.1088/1475-7516/2021/05/057}

\bibitem[{Tycner {et~al.}(2006)Tycner, Gilbreath, Zavala, Armstrong, Benson, Hajian, Hutter, Jones, Pauls, \& White}]{Tycner:2006nw}
Tycner, C., Gilbreath, G.~C., Zavala, R.~T., {et~al.} 2006, Astron. J., 131, 2710, \dodoi{10.1086/502679}

\bibitem[{{van Leeuwen}(2007)}]{V07}
{van Leeuwen}, F. 2007, \aap, 474, 653, \dodoi{10.1051/0004-6361:20078357}

\bibitem[{Virtanen {et~al.}(2020)Virtanen, Gommers, Oliphant, Haberland, Reddy, Cournapeau, Burovski, Peterson, Weckesser, Bright, {van der Walt}, Brett, Wilson, Millman, Mayorov, Nelson, Jones, Kern, Larson, Carey, Polat, Feng, Moore, {VanderPlas}, Laxalde, Perktold, Cimrman, Henriksen, Quintero, Harris, Archibald, Ribeiro, Pedregosa, {van Mulbregt}, \& {SciPy 1.0 Contributors}}]{2020SciPy-NMeth}
Virtanen, P., Gommers, R., Oliphant, T.~E., {et~al.} 2020, Nature Methods, 17, 261, \dodoi{10.1038/s41592-019-0686-2}

\bibitem[{Zmija {et~al.}(2023)Zmija, Vogel, Wohlleben, Anton, Zink, \& Funk}]{10.1093/mnras/stad3676}
Zmija, A., Vogel, N., Wohlleben, F., {et~al.} 2023, Monthly Notices of the Royal Astronomical Society, 527, 12243, \dodoi{10.1093/mnras/stad3676}

\end{thebibliography}
\bibliographystyle{aasjournal}
\clearpage

\end{document}